\documentclass[article]{aa}

\usepackage{natbib}
\usepackage{graphics,graphicx}
\usepackage{amsmath, amssymb}
\usepackage{setspace}
\usepackage{multirow}
\usepackage{color}
\usepackage{subcaption}
\usepackage{placeins}

\newcommand{\msol}{\ensuremath{\mathrm{M}_{\odot}}}

\newcommand{\mgas}{\ensuremath{M_{\rm gas}}}
\newcommand{\mdust}{\ensuremath{M_{\rm dust}}}
\newcommand{\lsol}{\ensuremath{\mathrm{L}_{\odot}}}

\newcommand{\lstar}{\ensuremath{L_{\star}}}

\newcommand{\tkin}{\ensuremath{T_{\textrm{kin}}}}

\newcommand{\mjup}{M$_{\rm Jup}$}

\newcommand{\gdrat}{\ensuremath{\Delta_{\rm g/d}}}

\begin{document}

\title{Mass constraints for 15 protoplanetary disks from HD\,$1$\,--\,$0$}
%\author{tbd}
%\author{M. Kama, L. Trapman, D. Fedele, S. Bruderer, E.F. van Dishoeck, M. Hogerheijde, E.A. Bergin, A. Miotello}

   \author{ M. Kama\inst{1,2}
            \and 
            L. Trapman\inst{3}
            \and
            D. Fedele\inst{4}
            \and 
            S. Bruderer\inst{5}
            \and
            M.R. Hogerheijde\inst{2,6}
            \and 
            A. Miotello\inst{7}
            \and 
            E.F. van Dishoeck\inst{2,5}
            \and
            C. Clarke\inst{1}
            \and 
            E.A. Bergin\inst{8}
          }

    \institute{
            Institute of Astronomy, University of Cambridge, Madingley Road, Cambridge CB3 0HA, UK \\
            \email{mkama@ast.cam.ac.uk}
        \and
            Tartu Observatory, Observatooriumi 1, T\~{o}ravere 61602, Tartu, Estonia
        \and
            Leiden Observatory, Leiden University, Niels Bohrweg 2, NL-2333 CA Leiden, The Netherlands
        \and
            INAF–Osservatorio Astrofisico di Arcetri, Largo E. Fermi 5, 50125 Firenze, Italy
        \and
            Max-Planck-institute f\"{u}r extraterrestrische Physic, Giessenbachstra{\ss}e, D-85748 Garching bei M\"unchen, Germany
        \and
            Anton Pannekoek Institute for Astronomy, University of Amsterdam, Science Park 904, 1090 GE Amsterdam, The Netherlands
        \and
            European Southern Observatory, Karl-Schwarzschild-Str. 2, D-85748 Garching bei M\"unchen, Germany
        \and
            Department of Astronomy, University of Michigan, 1085 S. University Ave, Ann Arbor, MI 48109
             }

% 5 {} token are mandatory
  \abstract
  %Protoplanetary disk gas mass measurements are difficult. 
   {Hydrogen deuteride (HD) rotational line emission can provide reliable protoplanetary disk gas mass measurements, but it is difficult to observe and detections have been limited to three T-Tauri disks. No new data have been available since the \emph{Herschel} Space Observatory mission ended in 2013.}
   {We set out to obtain new disk gas mass constraints by analysing upper limits on HD\,$1$\,--\,$0$ emission in \emph{Herschel}/PACS archival data from the DIGIT key programme.}
   {With a focus on the Herbig Ae/Be disks, whose stars are more luminous than T Tauris, we determine upper limits for HD in data previosly analysed for its line detections. Their significance is studied with a grid of models run with the DALI physical-chemical code, customised to include deuterium chemistry.}
   {Nearly all the disks are constrained to $M_{\rm gas}\leq0.1\,$M$_{\odot}$, ruling out global gravitational instability. A strong constraint is obtained for the HD 163296 disk mass, $M_{\rm gas} \leq 0.067\,$M$_{\odot}$, implying $\gdrat\leq100$. This HD-based mass limit is towards the low end of CO-based mass estimates for the disk, highlighting the large uncertainty in using only CO and suggesting that gas-phase CO depletion in HD 163296 is at most a factor of a few. The \mgas\ limits for HD 163296 and HD 100546, both bright disks with massive candidate protoplanetary systems, suggest disk-to-planet mass conversion efficiencies of $M_{\rm p}/(\mgas + M_{\rm p})\approx10$ to $40\,$\% for present-day values. Near-future observations with SOFIA/HIRMES will be able to detect HD in the brightest Herbig~Ae/Be disks within $150\,$pc with $\approx10\,$h integration time.} 
  {}
%  {\lt{Rotational lines of HD remain one of the most precise methods to measure the gas mass of protoplanetary disks.}% and the higher temperatures of Herbig disks makes them ideal targets for future HD observations.}

\maketitle

\section{Introduction}
\label{sec: introduction}

%The total gas mass (\mgas) of a protoplanetary disk controls planet formation after the cloud collapse stage.  It is also important for absolute gas-phase chemical abundance studies, and factors into the Toomre $Q$ parameter which quantifies whether a disk is gravitationally stable or not \citep{Toomre1964}.  We analyse \emph{Herschel} archival data to obtain new constraints on \mgas\ for a sample of Herbig~Ae/Be disks, and narrow down the mass of the HD\,163296 disk to a factor of a few.

%The total gas mass of a protoplanetary disk, \mgas,  is an elusive quantity, robustly measured in only a handful of cases.  
The elusive total gas mass of a protoplanetary disk is relevant for planet formation, dust dynamics, and for testing disk evolution models.  Due to difficulties in observing H$_{2}$, \mgas\ has been robustly measured in only three cases \citep{Berginetal2013, McClureetal2016}.  In this work, we use \emph{Herschel} archival data to constrain \mgas\ in a sample of $15$ Herbig~Ae/Be disks, and determine the mass of HD\,163296 to within a factor of a few.

%H$_{2}$ rotational transitions in the mid-infrared, probing gas at $\sim 100$ to $\sim 1000\,$K, have yielded detections of hot gas in some disks \citep{Bitneretal2007, MartinZaidietal2007, Carmonaetal2008b, Bitneretal2008}. Detections of H$_{2}$ electronic transitions in the ultraviolet are of use as line-of-sight probes.
%H$_{2}$ has no permanent dipole moment so the total gas mass cannot be easily measured, although FUV, NIR, and MIR observations of H$_{2}$ emission and absorption lines have yielded some constraints.
The gas mass is dominated by H$_{2}$, which has a large energy spacing between its lowest rotational levels (\emph{para}-H$_{2}\,J=2$\,--\,$0$, $\Delta E=512\,$K) and lacks a dipole moment.  As such, H$_{2}$ is not emissive at the $10$-$100\,$K temperatures typical for disks.  Dust continuum emission at millimetre wavelengths is often used to estimate \mgas. Gas and dust are linked through a mass ratio, canonically \gdrat~$=100$ for solar-composition material below $\sim 10^{3}\,$K \citep[e.g.][]{Lodders2003}.  While dust emission is easy to detect, the different dust and gas evolution as well as uncertain opacity values limit its reliability in measuring \mgas.  The most precise \mgas\ measurements to-date are from hydrogen deuteride (HD) rotational lines.  The relative abundance of this deuterated isotopolog of H$_{2}$ is set by the local absolute atomic ratio, D/H~$=(2.0\pm0.1)\times10^{-5}$ \citep[][]{Prodanovicetal2010}, and is minimally affected by disk chemistry \citep{Trapmanetal2017}.  As the $J=1$ rotational level is at $E/k_B =128.5\,$K, HD emits from warm gas \citep[$T_{\rm gas}\approx30$ to $50\,$K,][]{Berginetal2013,Trapmanetal2017}.  This is sufficient to constrain the total \mgas, especially if the temperature structure is constrained via other observables.  The HD\,$J=1$\,--\,$0$ line at $112\,\mu$m, is however impossible to observe from the ground due to atmospheric absorption and requires air- or spaceborne telescopes.

After the pioneering HD\,$1$\,--\,$0$ detection in TW\,Hya \citep{Berginetal2013, Trapmanetal2017}, facilitated by the PACS spectrometer \citep{Poglitschetal2010} on the \emph{Herschel Space Observatory} \citep{Pilbrattetal2010}, further detections were only made in DM\,Tau and GM\,Aur \citep{McClureetal2016} before the instrument expired.  The masses of these T~Tauri disks are $\mgas=(6-9)\times10^{-3}$, $(1-4.7)\times10^{-2}$, and $(2.5-20.4)\times10^{-2}\,$\msol, respectively.  An upper limit $\mgas\leq8\times10^{-2}\,$\msol\ was obtained for the Herbig~Ae/Be system HD\,100546 \citep[][revised down from the published value due to a mistake in the D/H ratio]{Kamaetal2016b}. 
%All three sources have a disk-averaged gas-to-dust ratio $\gdrat\approx 100$.  
%With \emph{Herschel} no longer available, there are currently no facilities capable of observing HD. 

In this work, we use the 2D physical-chemical code \texttt{DALI} \citep{Brudereretal2012,Brudereretal2013} to constrain \mgas\ in $15$ disks by analysing \emph{Herschel} archival data covering the HD\,$1$\,--\,$0$ and $2$\,--\,$1$ lines.  The data and models are discussed in Sections \ref{sec: observations and sample} and \ref{sec: modelling}, respectively. In Section \ref{sec: results}, we explore the disk mass constraints, with a focus on HD\,163296, and discuss the potential for gravitational instability.  In Section~\ref{sec:discussion}, we compare the mass of disks, stars, and planetary systems for stars over $1.4\,$\msol.  We also discuss future observations of HD with SOFIA/HIRMES \citep{Richardsetal2018} and SPICA/SAFARI \citep{Nakagawaetal2014, Audleyetal2018}.

%coming HIRMES instrument on SOFIA, due for commisioning in 2019 \citep[e.g.][]{Richardsetal2018}, and future far infrared space-based mission such as the proposed SPICA (e.g., \citealt{Nakagawaetal2014}) and OST (e.g., \citealt{Bonatoetal2019}) missions.

\section{Observations and sample}
\label{sec: observations and sample}

We use archival data from the \emph{Herschel Space Observatory} \citep{Pilbrattetal2010} key programme DIGIT (PI N.J. Evans), which targeted $30$ protoplanetary disks with the PACS \citep{Poglitschetal2010} instrument at $50$--$210\,\mu$m.  Detected gaseous species in this data were presented in \citet{Fedeleetal2013} and \citet{Meeusetal2013}.  We analyse upper limits on HD $J=1$\,--\,$0$ and $2$\,--\,$1$ lines at $112$ and $56\,\mu$m for the 15 Herbig Ae/Be disks in the sample.  Due to the intrinsically higher luminosity of their host stars ($\sim10\,$-$100\,$\lsol), these disks are warmer, and brighter in continuum and line emission than those around T~Tauri stars. This enables tighter constraints for disks at equivalent distance.  

We selected disks around stars of spectral type mid-F to late-B, including well-known targets such as HD\,100546 and HD\,163296.  HD\,50138 was excluded as it is likely an evolved star \citep{Ellerbroeketal2015}, and HD\,35187 because it is a binary of two intermediate-mass stars and not directly comparable to our model grid.  The data are spectrally unresolved, with $\delta v\approx100\,$km$\,$s$^{-1}$ ($\lambda/\delta\lambda=3000$) at the shortest wavelengths ($51\,\mu$m), while expected linewidths are $\leq10\,$km$\,$s$^{-1}$.  Exposure times ranged from $4356\,$s to $8884\,$s.  The system parameters and $3\sigma$ line flux upper limits are given in Table~\ref{tab:obs}. 

%We draw upon the DIGIT data as reported by \citet{Fedeleetal2013}, selecting only the Herbig~Ae/Be disks for which the $112\,\mu$m and $56\,\mu$m wavelengths were covered ($50$--$73$ and $100$--$145\,\mu$m spectral ranges). 

We obtained flux limits for the HD transitions from the $1\sigma$ noise reported for the nearest lines of other molecules from \citet{Fedeleetal2013}:  OH~$^{2}\Pi_{1/2}\,J=9/2^{-}$--$7/2^{+}$ at $55.89\,\mu$m for the $56\,\mu$m line and OH~$^{2}\Pi_{3/2}\,J=5/2^{-}$--$3/2^{+}$ at $119.23\,\mu$m for the $112\,\mu$m line.  With a typical $1\sigma$ uncertainty of $5\times10^{-18}\,$W\,m$^{-2}$ at $112\,\mu$m and $2\times10^{-17}$W\,m$^{-2}$ at $56\,\mu$m, neither of the HD lines is detected in the targets, individually or stacked.  For comparison, the HD\,$1$\,--\,$0$ detections \citet{Berginetal2013} and \citet{McClureetal2016} had respective uncertainties of roughly $7\times10^{-19}\,$W\,m$^{-2}$ and $5\times10^{-19}\,$W\,m$^{-2}$, which illustrates the difference between those targeted, deep integrations and the survey-type observations analysed here.

%Stacking the spectra for all the sources did not yield a detection.
%OH $^{2}\Pi_{1/2}\,9/2^{-}$--$7/2^{+}$ at $55.89\,\mu$m for the $56\,\mu$m line and CO $J$\,=\,23\,-\,22 at $113.5\,\mu$m for the $112\,\mu$m line. Stacking the spectra for all the sources did not yield a detection.

%The OH and HD 1-0 line are further apart, but the spectra do not show any significant variation in noise levels between 110 and 120 micron.

The disks fall into two categories, cold \citep[flat, group~II in the Meeus classification,][]{Meeusetal2001} and warm (flaring, group~I). This characterises the shape of the radial optically thick surface, where starlight is effectively absorbed.  Starlight impinges at a shallow angle on flat disks, and heating is inefficient compared to that above the same midplane location in a flaring disk.  In addition, among the Herbig~Ae/Be systems flaring, group~I disks have resolved cavities or gaps $10$--$100\,$au scales in their millimetre dust emission \citep{Maaskantetal2013, Kamaetal2015}.

\begin{table*}
\centering
\caption{HD line flux upper limits ($3\sigma$) for the sample.}
\begin{tabular}{ c c c c c c c c }
\hline\hline
Name & $L_{\star}$ & $T_{\rm eff}$ & d  & HD~$112\,\mu$m & HD~$56\,\mu$m & $F_{\rm 1.3mm}$ & Meeus \\
	& (L$_{\odot}$) & (K) & (pc) & $\left(10^{-17}\frac{\rm W}{\rm m^{2}}\right)$ & $\left(10^{-17}\frac{\rm W}{\rm m^{2}}\right)$ & (mJy) & group \\
\hline
HD 104237	& $26^{\rm F15}$ & $8000^{\rm F15}$ & $108^{\rm GDR2}$ & $\leq 0.9$ & $\leq 2.4$ & $92\pm19^{\rm M14}$ & IIa \\
HD 144668	& $58^{\rm F15}$ & $8500^{\rm F15}$ & $161^{\rm GDR2}$ & $\leq 0.8$ & $\leq 7.8$ & $20\pm16^{\rm M14}$ & IIa \\
HD 163296	& $31^{\rm F12}$ & $9200^{\rm F12}$ & $101^{\rm GDR2}$ & $\leq 0.6$ & $\leq 3.0$ & $743\pm15^{\rm M14}$ & IIa \\
\hline
HD 31293	& $59^{\rm F15}$ & $9800^{\rm F12}$ & $139^{\rm F12}$ & $\leq 4.2$ & $\leq 22.4$ & $136\pm15^{\rm M14}$ & Ia \\
HD 36112	& $22^{\rm M14}$ & $8190^{\rm F12}$ & $160^{\rm GDR2}$ & $\leq 0.6$ & $\leq 7.6$ & $72\pm13^{\rm M14}$ & Ia \\
HD 38120	& $123^{\rm S13}$ & $10471^{\rm S13}$ & $406^{\rm GDR2}$ & $\leq 0.9$ & $\leq 5.6$ & - & Ia \\
HD 100546	& $36^{\rm K16b}$ & $10390^{\rm K16b}$ & $110^{\rm GDR2}$ & $\leq 2.7$ & $\leq 16.0$ & $465\pm20^{\rm M14}$ & Ia \\
HD 139614	& $6.6^{\rm F15}$ & $7750^{\rm F15}$ & $135^{\rm GDR2}$ & $\leq 1.2$ & $\leq 8.5$ & $242\pm15^{\rm M14}$ & Ia \\
HD 142527	& $7.9^{\rm F15}$ & $6500^{\rm F15}$ & $157^{\rm GDR2}$ & $\leq 4.0$ & $\leq 13.0$ & $1190\pm33^{\rm M14}$ & Ia \\
HD 179218	& $110^{\rm F12}$ & $9640^{\rm F12}$ & $266^{\rm GDR2}$ & $\leq 1.1$ & $\leq 7.0$ & $71\pm7^{\rm M14}$ & Ia \\
\hline
HD 97048	& $33^{\rm F15}$ & $10500^{\rm F15}$ & $171^{\rm F15}$ & $\leq 2.4$ & $\leq 2.4$ & $454\pm34^{\rm M14}$ & Ib \\
HD 100453	& $8.5^{\rm F15}$ & $7250^{\rm F15}$ & $104^{\rm GDR2}$ & $\leq 1.3$ & $\leq 5.5$ & $200\pm21^{\rm M14}$ & Ib \\
HD 135344B	& $7.1^{\rm F15}$ & $6375^{\rm F15}$ & $136^{\rm GDR2}$ & $\leq 0.6$ & $\leq 8.2$ & $142\pm19^{\rm M14}$ & Ib \\
HD 169142	& $10^{\rm F12}$ & $7500^{\rm F12}$ & $114^{\rm GDR2}$  & $\leq 2.4$ & $\leq 13.5$ & $197\pm15^{\rm M14}$ & Ib \\
Oph IRS 48$^{\star}$	& $14.3^{\rm S13}$ & $9000^{\rm S13}$ & $134^{\rm GDR2}$ & $\leq 1.2$ & $\leq 8.3$ & $60\pm10^{\rm M14}$ & Ib \\
\hline\\
\end{tabular}
\newline
\begin{flushleft}
\emph{Notes: }$^{\star}$ -- WLY\,2-48.\\
\emph{References:} F12 -- \cite{Folsometal2012}; S13 -- \cite{Salyketal2013}; M14 -- \cite{Maaskantetal2014} and references therein; F15 -- \cite{Fairlambetal2015}; K16b -- \cite{Kamaetal2016b}; GDR2 -- \cite{Brownetal2018}. 
\end{flushleft}
\label{tab:obs}
\end{table*}

\section{Modelling}
\label{sec: modelling}

\subsection{DALI}
\label{sec: DALI}

To determine the behaviour of the HD\,$1$\,--\,$0$ line and $1.3\,$millimetre continuum flux as a function of disk structure parameters, we run a grid of models with the 2D physical-chemical disk code DALI \citep{Brudereretal2012,Brudereretal2013}. The surface density is parameterized following the viscous accretion disk formalism \citep{Lynden-BellPringleetal1974,Hartmann1998}:
%The surface density a tapered powerlaw is assumed motivated by a disk structure that is set by viscous accretion, where $\nu \propto R^{\gamma}$ \citep{Lynden-BellPringleetal1974,Hartmann1998}.

\begin{equation}
\label{eq: surface density}
\Sigma_{\rm gas} = \Sigma_c \left( \frac{R}{R_c} \right)^{\gamma} \exp\left[ -\left(\frac{R}{R_c}\right)^{2-\gamma}\right],
\end{equation}
where $\Sigma_{c}$ is the surface density at the characteristic radius $R_{c}$, and $\gamma$ the power-law index which is generally $1$. Assuming an isothermal structure in hydrostatic equilibrium, the vertical structure is given by a Gaussian density distribution \citep{KenyonHartmannetal1987}:

\begin{equation}
\label{eq: gas density}
\rho_{\rm gas}(R,z) = \frac{\Sigma_{\rm gas}(R)}{\sqrt{2\pi}Rh}\exp\left[-\frac{1}{2}\left(\frac{z}{Rh} \right)^2 \right].
\end{equation}
Here $h = h_c (R/R_c)^{\psi}$, $\psi$ is the flaring index and $h_c$ is the disk opening angle at $R_c$.

%The dust grains in the disk are split into two populations in simulate to include dust settling. 
A population of small grains (0.005-1 $\mu$m), with a mass fraction $f_{\rm small}$, follows the gas density distribution given in Eq.~\eqref{eq: gas density}. A second population, consisting of large grains (1 $\mu$m - 1 mm), has a mass fraction $f_{\rm large}$. Their scale height is $\chi h$, where $\chi\in{(0,1]}$ is the settling parameter.

For the dust opacities of both small and large grain populations we assume a standard interstellar composition following \citet{WeingartnerDraine2001}, in line with \citet{Brudereretal2013}.  The absorption coefficient for the small (large) grains is $29.9\,$cm$^{2}\,$g$^{-1}$ ($30.0\,$cm$^{2}\,$g$^{-1}$) at $112\,\mu$m and $154\,$cm$^{2}\,$g$^{-1}$ ($46.3\,$cm$^{2}\,$g$^{-1}$) at $56\,\mu$m.

First, the radiation field and dust temperature are determined from Monte Carlo radiative transfer. Next, the gas temperature (heating-cooling balance) and chemical composition (steady-state) are solved for iteratively.  Raytracing then yields simulated line and continuum observations.
%: For each location in the disk the time-dependent chemistry is solved to determine the abundances of molecular and atomic species. Next the excitation levels of the species are calculated assuming non-LTE. Based on the excitation levels the gas temperature is computed by balancing the heating and cooling processes. The resulting gas temperature is used to re-evalute the abundances and excitation levels, until a self-consistent solution is found.

\subsubsection{HD chemical network \emph{versus} fixed abundance}
\label{sec: HD abundances}
The HD abundance (HD/H$_{2}$) can be prescribed as a constant or obtained from solving a chemical reaction network. 

In the parametric approach, the HD abundance is determined by the local D/H ratio, which for the local ISM (within $\approx 2\,$kpc) is measured to be (D/H)$_{\rm ISM} = (2.0\pm0.1)\times10^{-5}$ \citep{Prodanovicetal2010}. Assuming all deuterium is in HD, this gives HD/H$_{2}=4\times10^{-5}$. 

A more refined approach is to calculate the HD abundance using a reaction network which includes deuterium. \citet{Trapmanetal2017} extended the standard DALI chemical network \citep[originally based on the UMIST06 database][]{Woodalletal2007} to include the species HD, D, HD$^+$, and D$^+$. HD formation on dust and ion-exchange reactions were included, in addition to HD self-shielding. The details of the implementation are described in Section 2.3 of \cite{Trapmanetal2017}. 

Using the chemical network approach, we find that all of the available deuterium is locked up in HD for the vast majority of the disk, and the parametric abundance of HD/H$_{2}=4\times10^{-5}$ is appropriate to use. The network produces less HD in only two regions: the uppermost layers of the disk where HD is photodissociated, and in a thin intermediate layer, where the HD abundance is decreased by a factor of $\sim2$. Tests determined that neither of these significantly affects the disk-integrated HD line flux.

Given the very close match between the two approaches, we opt for simplicity and fix the HD/H$_{2}$ ratio at $4\times 10^{-5}$.

\begin{table}[htb]
  \centering   
  \caption{\label{tab: grid parameters}\texttt{DALI} model grid parameters.}
  %\begin{tabular*}{0.95\columnwidth}{l c}
  \begin{tabular}{l c}
    \hline\hline
    Parameter & Range\\
    \hline
     \multicolumn{2}{c}{\emph{Chemistry}}\\
     \hline
     Chemical age & 1 Myr\\
     HD/H$_{2}$ & $4\cdot10^{-5}$ \vspace{0.1cm} \\
     \hline
     \multicolumn{2}{c}{\emph{Physical structure}}\\
     \hline
     $\gamma$ &  1.0\\ 
     $\psi$ & [0.0, 0.3]\\ 
     $h_{\rm c}$ &  [0.05, 0.15] rad\\ 
     $R_{\rm c}$ & [50, 150] au \\ 
     $M_{\rm gas}$ & $[10^{-3}, 10^{-2}, 10^{-1}]$ M$_{\odot}$ \vspace{0.1cm} \\
     \hline
     \multicolumn{2}{c}{\emph{Dust properties}}\\
     \hline
     \gdrat & [10, 50, 100, 300] \\
     %Gas-to-dust ratio & [10, 50, 100, 300] \\
     $f_{\rm large}$ & [0.8, 0.95] \\
     $\chi$ & [0.2, 0.5] \\
     $f_{\rm PAH}$ & 0.001 \vspace{0.1cm} \\
     \hline
     \multicolumn{2}{c}{\emph{Stellar properties$^1$}}\\
     \hline
     %type & Herbig \lt{(SpType B9.5V)} \\
     $T_{\rm eff}$ & 10390 K\\
     $L_{\rm X}$ & 8$\cdot10^{28}$ erg s$^{-1}$ \\
     $T_{\rm X}$ & 7$\cdot10^{7}$ K \\
     $L_{*}$ & [10, 50, 115] L$_{\odot}$ \\
     $\zeta_{\rm cr}$ & $10^{-17}\ \mathrm{s}^{-1}$ \vspace{0.1cm} \\
     \hline
     \multicolumn{2}{c}{\emph{Observational geometry}}\\
     \hline
     $i$ & 60$^{\circ}$\\
     d & 150 pc\\
    \hline
  \end{tabular}
  %\end{tabular*}
  %\captionsetup{width=.9\columnwidth}
  %\lt{See Section \ref{sec: DALI} for definitions of symbols.} \\ %and Bruderer+2012
  \flushleft
  \emph{Notes: }Standard \texttt{DALI} parameter names as in \citet{Brudereretal2012}. Deuterium abundance from \citet{Prodanovicetal2010}. $^1$HD\,100546 \citep{Brudereretal2012}. 
\end{table}

Part of our analysis below involves modelling CO rotational lines. Due to processes such as chemical conversion and freeze-out, the gas-phase total abundance of C and O nuclei can be more than a factor of ten below nominal \citep[e.g.][]{Favreetal2013, Kamaetal2016b}, which makes CO-based mass estimates highly uncertain.  We refer to the reduction of gas-phase C and O nuclei below their total values with the term \emph{depletion}, and the phenomenon can be included in our modelling as a reduction of the total amount of volatile C or O input into a given DALI model. This is relevant for Section\,\ref{sec:hd163296}, in particular.

\subsection{Model grid}

\begin{figure}[!ht]
\includegraphics[clip,trim={1.3cm 0.6cm 1.85cm 1.3cm},width=1.0\columnwidth]{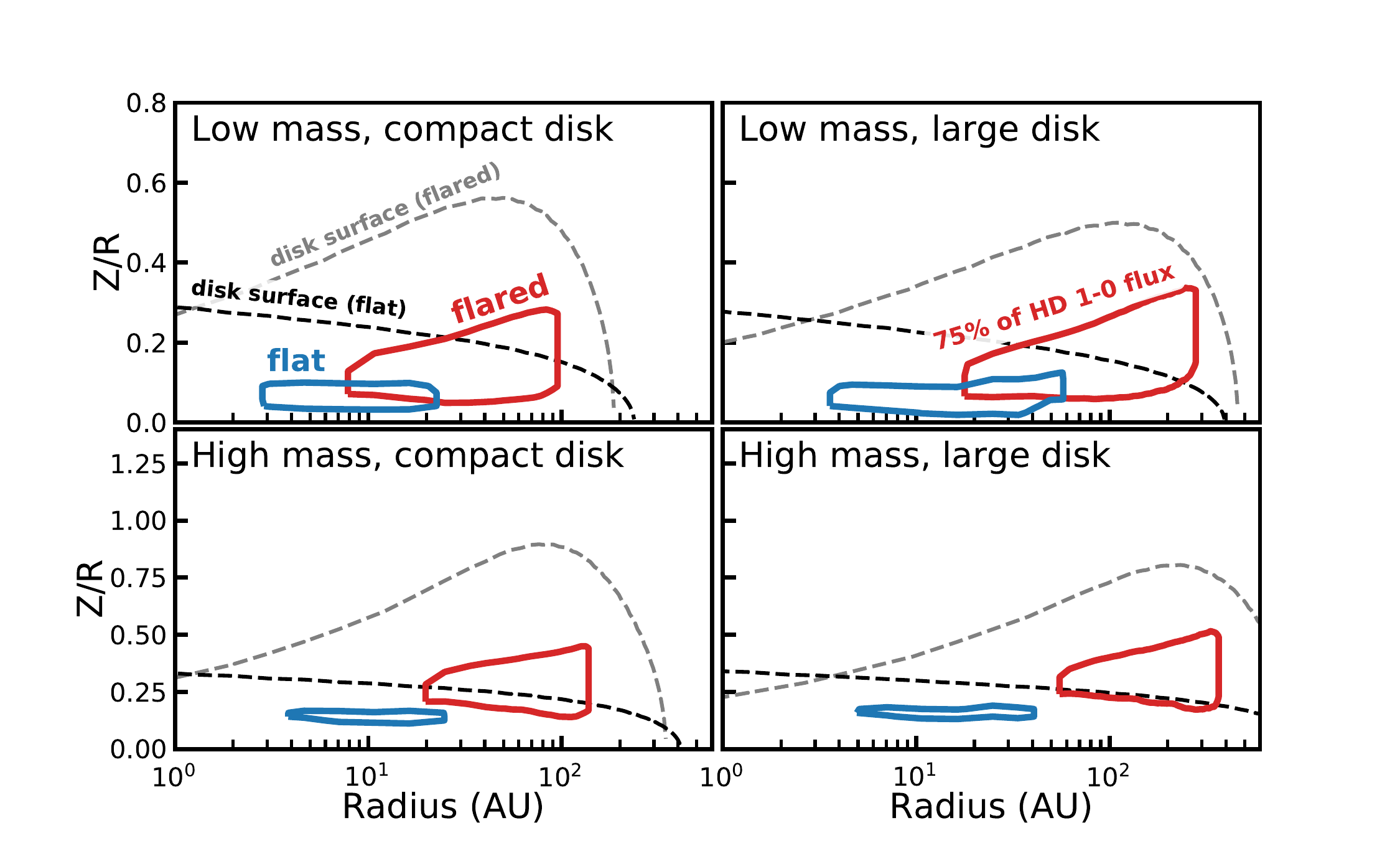}
% \begin{subfigure}{\columnwidth}
% \includegraphics[clip,trim={1.3cm 0.6cm 1.85cm 1.3cm},width=1.0\columnwidth]{HD_10_cbfs_flat}
% \end{subfigure}
% \begin{subfigure}{\columnwidth}
% \includegraphics[clip,trim={1.3cm 0.6cm 1.85cm 1.3cm},width=1.0\columnwidth]{HD_10_cbfs_flared}
% \end{subfigure}
\caption{HD\,$1$--$0$ line emitting regions in our flat/cold (blue) and flared/warm (red) disk models.  Solid contours contain the middle $75$\% of vertically cumulative line emission.  Dashed lines are gas number density iso-contours for $n_{\rm gas} = 10^6\ \mathrm{cm}^{-3}$, acting as a disk ``outline''.}
% \caption{HD $1$--$0$ line emitting regions in our flat/cold (blue, top panels) and flared/warm (red, bottom panels) disk models. The solid contours denote radial and vertical cumulative emission reaching $25$ and $75$\%. The dashed lines show where the disk density equals $n_{\rm gas} = 10^6\ \mathrm{cm}^{-3}$. 
% \lt{split figures or combined figures??}

\label{fig: hdcbfs}
\end{figure}

\begin{figure*}[!ht]
\centering
\begin{subfigure}{0.99\linewidth}
\includegraphics[clip=,width=1.0\linewidth]{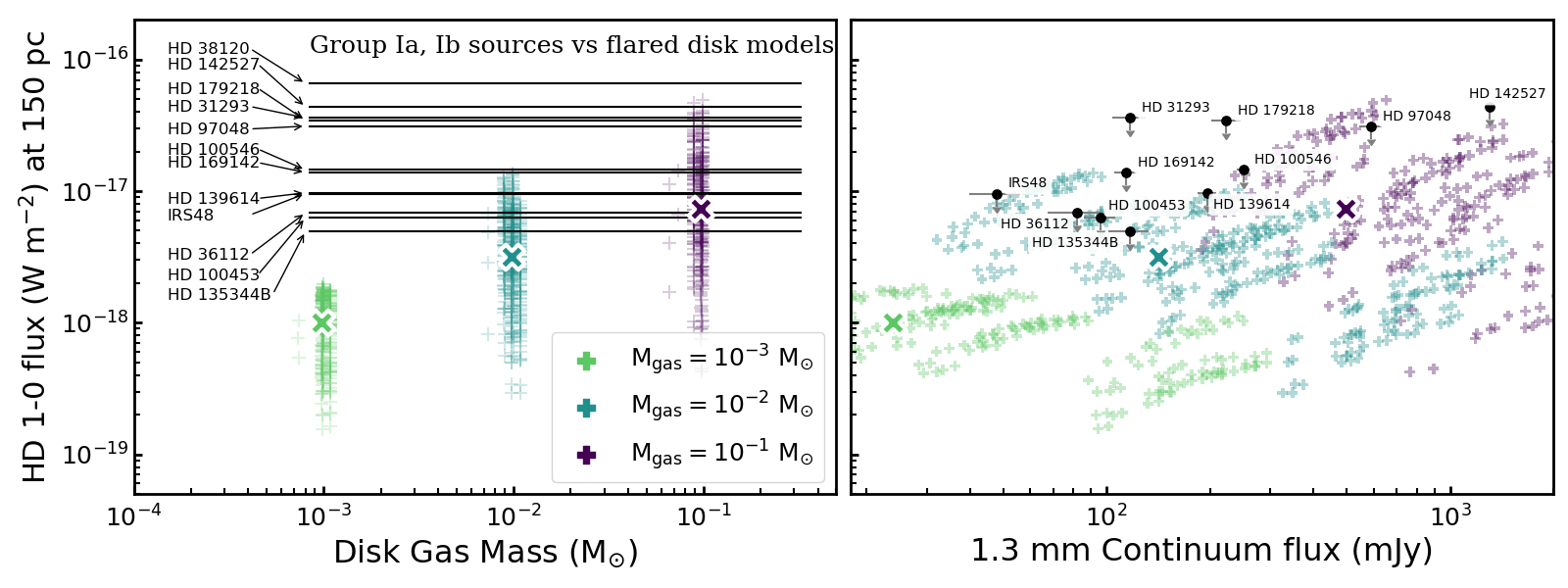}
\end{subfigure}
\begin{subfigure}{0.99\linewidth}
\includegraphics[clip=,width=1.0\linewidth]{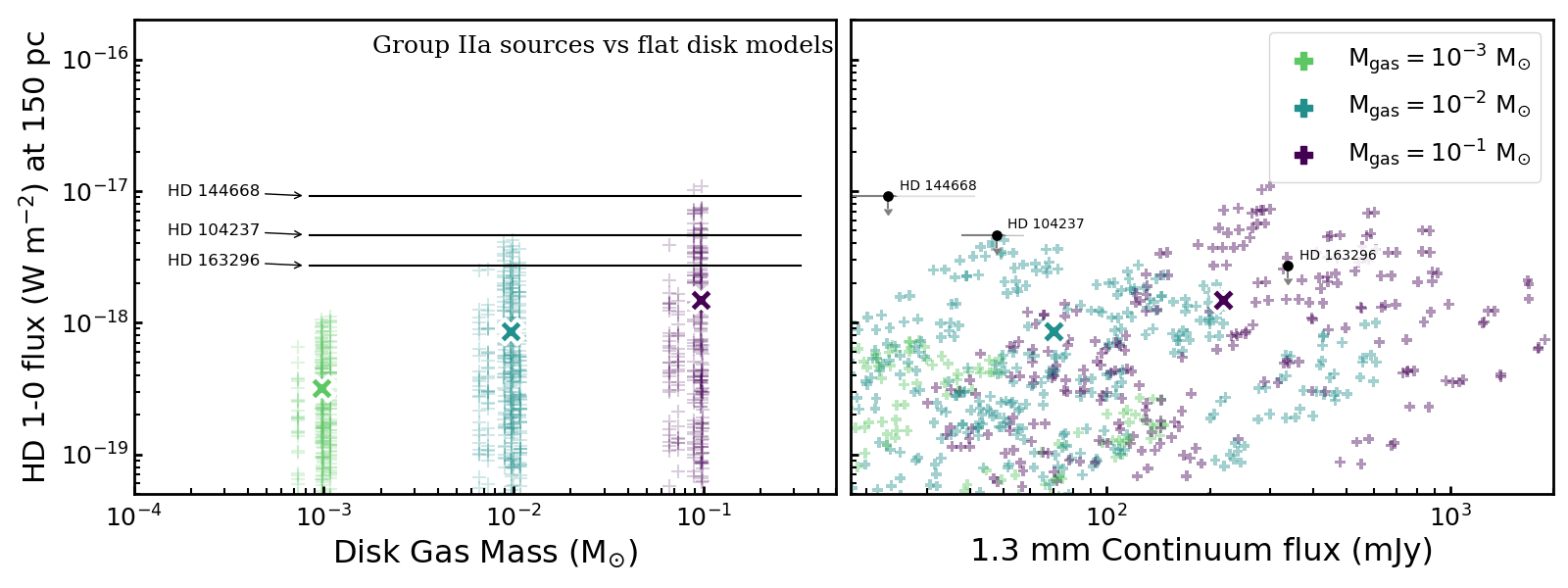}
\end{subfigure}
\caption{Distance-normalised $3\sigma$ upper limits on HD~$112\,\mu$m line flux for the disk sample (black lines and circles) compared with our grid of DALI disk models (coloured crosses).  Highlighted crosses show the HD~$112\,\mu$m line flux of our fiducial model.  The top panels show the group I sources compared to models with flaring angle $\psi = 0.3$. The bottom panels show the group II sources compared to models with $\psi = 0.0$. \textbf{Left:} models are separated based on gas mass. \textbf{Right:} HD\,$1$\,--\,$0$ upper limits set against 1.3 mm continuum fluxes for both observations and models.
}
\label{fig:HD10_Mgas_1mm}
\end{figure*}

To investigate the range of disk properties constrained by the \emph{Herschel} upper limits on the HD\,$1$\,--\,$0$ line, we run a grid of Herbig~Ae/Be disk models covering a wide range of parameters, summarized in Table \ref{tab: grid parameters}. The disk gas masses are $M_{\rm gas} = 10^{-3}$, $10^{-2}$, and $10^{-1}\,$M$_{\odot}$. Dust mass is defined by the gas-to-dust mass ratio, with values \gdrat~$=10$, $50$, $100$, and $300$, and ranges from $M_{\rm dust} = 3\times10^{-6}$ to $10^{-2}\,$M$_{\odot}$. The shape of the stellar spectrum, including UV excess, is based on HD\,100564 from \citet{Brudereretal2012}. The spectrum is scaled to the total stellar luminosity, \lstar~$\in [10,50,115]\,$L$_{\odot}$. This covers the sources in our sample, as given in Table~\ref{tab:obs}. In total we run 2304 models, with parameters given in Table~\ref{tab: grid parameters}.  Our fiducial model has $h_c=0.15$, $R_c=50\,$au, $\Delta_{\rm gd}=100$, $f_{\rm large}=0.95$, $\chi=0.2$, and $L_{\star}=10\,$\lsol.

%%%% OLD location Figure 2

%To give some reference for which parts of the disk emits in HD 1\,-\,0, 
Figure \ref{fig: hdcbfs} shows the HD\,$1$\,--\,$0$ emitting regions and disk mass outline for models representing extremes in flaring ($\Psi=0.0$ and $h_{\rm c}=0.05$ for flat, and $\Psi=0.3$ and $h_{\rm c}=0.15$ for flared), radial extent ($R_{\rm c}=50$ and $125\,$au), and total disk mass.  From the figure it is clear that the flared disk ($\psi=0.3,h_c=0.15$), shown in red, has a much large emitting region than the flat disk ($\Psi=0.0,h_c=0.05$), shown in blue. In both cases the HD\,$1$\,--\,$0$ emission originates from the warm layer above the midplane.

%\lt{-- We haven't introduced the fiducial model in the text (yet). This could be done either here or in the paragraph before that. --}

%The reference model is a flat (scaleheight exponent $\Psi=0$) disk of mass $0.01\,$\msol\ around a star with a luminosity of $10\,$\lsol. \edit{-- do we need this explanation here? it is not used anywhere (yet)--}

\section{Results}
\label{sec: results}

In Figure~\ref{fig:HD10_Mgas_1mm}, we show the HD $J$\,=\,1\,-\,0 flux as a function of \mgas\ and $1.3\,$millimetre continuum flux.  The warm, flaring, group~I disks and cold, flat, group~II disks are highlighted separately for clarity.

\subsection{Parameter dependencies in the grid}
%Several trends are evident in the models.  For a given \mgas, increasing the dust mass (decreasing the gas-to-dust mass ratio, \gdrat) decreases the HD flux, because the rising continuum opacity across the disk limits the mass of optically thin warm material.  One order of magnitude variation in stellar luminosity introduces a variation of roughly a factor of five in the continuum and HD emission. 

%\subsection{Parameter dependencies}
%\label{sec:analytical}
\begin{figure}[ht]
\centering
\includegraphics[clip=,width=1.0\columnwidth]{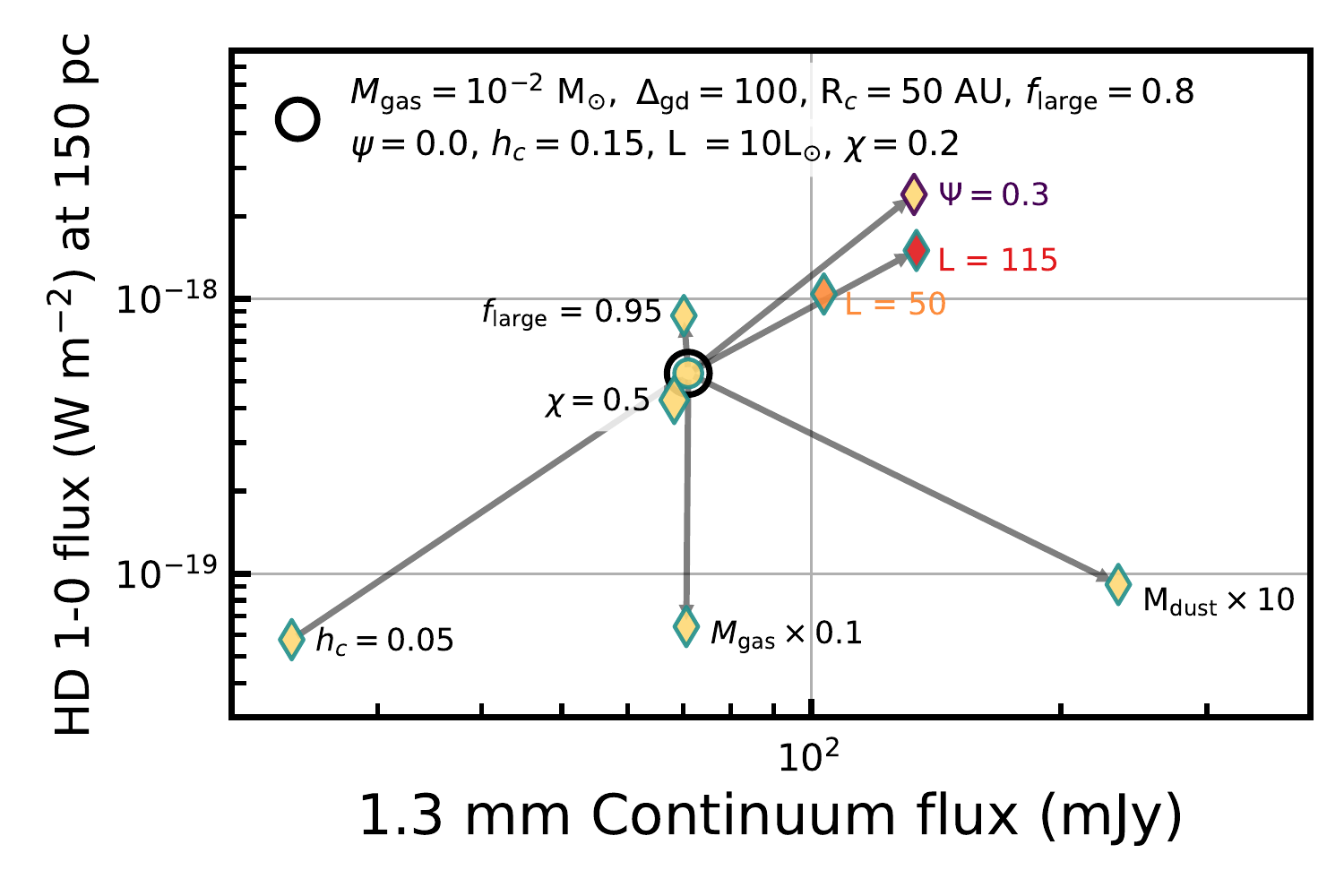}
\caption{HD\,$1$\,--\,$0$ line and dust continuum flux dependencies on disk and stellar parameters.}
\label{fig:analytical}
\end{figure}

Dependencies of the HD\,$1$\,--\,$0$ line and $1.3\,$millimetre continuum flux on the main model parameters are shown in Figure~\ref{fig:analytical}. The HD line flux depends linearly on \mgas, which has only a marginal effect on the dust emission. For a fixed \mgas, a $1\,$dex increase in \mdust\ leads to a factor $6.7$ lower HD and $2.5$ higher continuum flux. The flaring structure of the disk has the largest influence, as the HD line flux increases by a factor of $26$ when the flaring parameter $\Psi$ goes from $0$ (height is linear with radius, inefficient heating) to $0.3$ (very flared and efficiently heated). The Meeus group corresponds to the flaring structure (group~I disks are flared, II flat).

A near-linear dependence of HD line flux on \mgas\ arises because the HD line emission in the models is vertically limited by the dust optical depth $\tau$ at $112\,\mu$m out to $\approx 100\,$au radii, beyond which the surface density drops rapidly. Thus the HD contribution from the gas above and radially outside the dust scales linearly with the total gas mass. Dust emission, to first order, is optically thin at $1.3\,$mm, and thus scales linearly with the total dust mass. Again due to the dust optical depth dominating at the $112\,\mu$m wavelength of HD\,$1$\,--\,$0$, increasing the dust mass in a given column lifts the vertical $\tau(112\,\mu{\rm m})=1$ surface, hiding a larger fraction of the HD molecules.

\subsection{Constraints on \mgas\ across the sample}
\label{sec: gas mass constraints}

\begin{figure}[!ht]
\centering
\includegraphics[clip=,width=1.0\columnwidth]{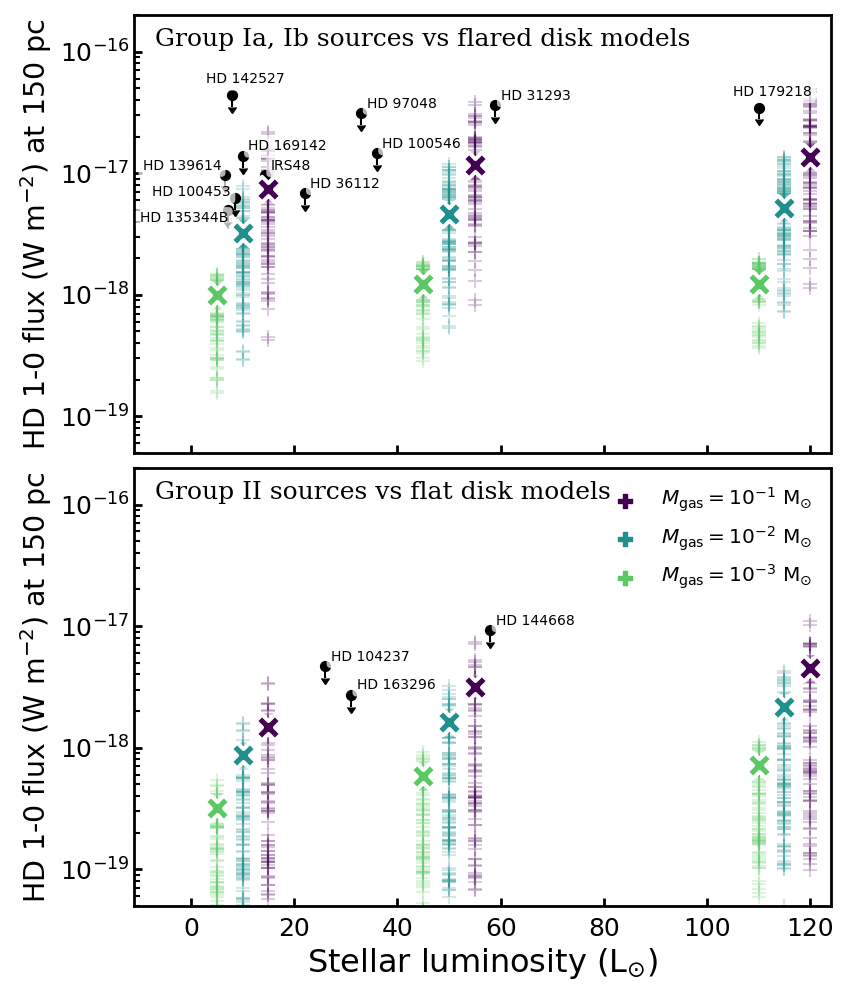}
\caption{HD\,$1$\,--\,$0$ line flux versus the stellar luminosity. Observed stellar luminosities taken from Table \ref{tab:obs}. Model stellar luminosities were given a small offset for clarity.  Highlighted crosses show our fiducial model ($h_c=0.15$, $R_c=50\,$au, $\Delta_{\rm gd}=100$, $f_{\rm large}=0.95$, $\chi=0.2$).}
\label{fig:HD_vs_Lstar}
\end{figure}

A comparison of the HD upper limits from \emph{Herschel} with our DALI model grid (Figures~\ref{fig:HD10_Mgas_1mm} and \ref{fig:HD_vs_Lstar}) places an upper limit of approximately $\mgas\leq0.1\,$\msol\ for the disks in our sample.  Among the flared, group~I disks (Fig.~\ref{fig:HD10_Mgas_1mm}, upper row), we find \mgas~$<0.02$--$0.03\,$\msol\ for IRS\,48, HD\,36112, HD\,100453, and HD\,135344B, while among the flat, group~II disks HD\,163296 has a limit at $<0.1\,$\msol. 

Source-specific models can tighten the mass limit for individual disks.  We run a small grid of models for HD\,163296, where we have a strong HD upper limit and a wide comparison range of indirect gas mass estimates from the literature based on various isotopologs of CO.
%and HD\,97048.  For HD\,139614 we currently lack the resolved CO observations of the outer disk needed to construct a source-specific model.
%note that this will change with Olja's ALMA data for this source.
%\lt{--update: we're not doing models for HD\,97048, and HD\,139614 is out of the blue now -- }

%HD\,163296 is flat, cold, group~II disk. It is conspicuously large, with a CO emission radius of $\approx 540\,$au \citep{Rosenfeldetal2013}.  HD\,97048 is strongly flared, warm, group~I disk \citep{Walshetal2016}. 

%%%%% OLD table/figure 163 location

\subsection{HD\,163296}\label{sec:hd163296}

\begin{figure}[!ht]
\centering
\includegraphics[clip=,width=1.0\columnwidth]{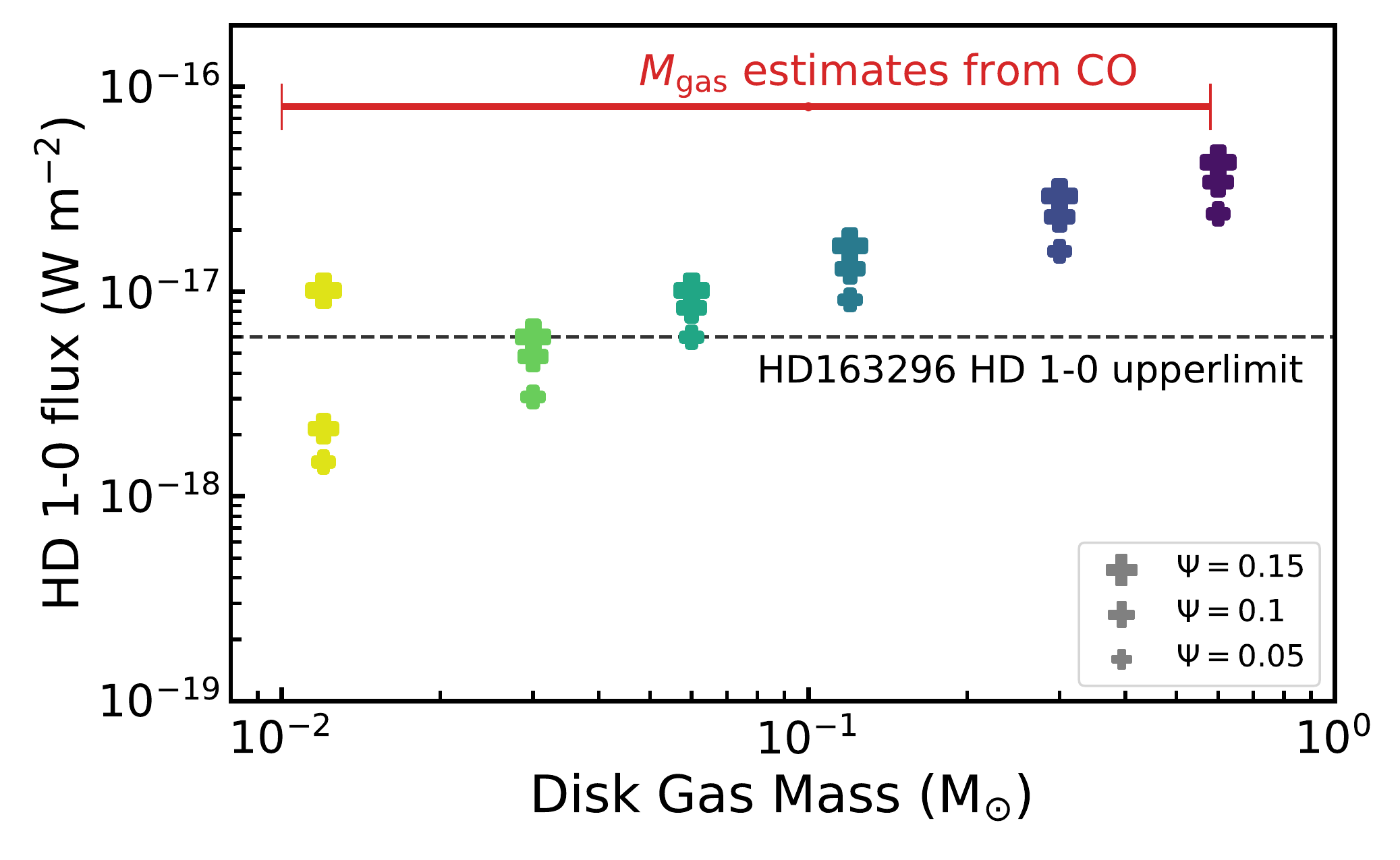}
\caption{\label{fig: HD163296 fit} Comparing the HD\,163296 specific models to the HD\,$1$\,--\,$0$ upper limit \citep{Fedeleetal2013}. All models have a dust mass $M_{\rm dust} = 6.6\times10^{-4}\ \mathrm{M}_{\odot}$ (Table~\ref{tab: parameters case studies}). The red bar shows the range of gas masses inferred from CO in the literature. }
\end{figure}

We constrain the gas mass in the HD\,163296 disk to \mgas~$\leq0.067\,$\msol\ (Figure~\ref{fig: HD163296 fit}).  Given that the disk-integrated dust mass in our model is $6.7\times 10^{-4}\,$\msol, this constrains the gas-to-dust ratio to \gdrat~$\leq100$ and has implications for the gas-phase volatile abundances, which we discuss below.  This source-specific model matches the continuum spectral energy distribution, $^{12}$CO rotational ladder and isotopolog lines, and several other key volatile species.  The full details of this modelling are outside the scope of this paper and will be published separately, below we focus on the main outcomes of the continuum, CO, and HD modelling.

\begin{table}[htb]
  \centering
  \caption{\label{tab: parameters case studies} Adopted model for HD\,163296}
  \noindent\makebox[\columnwidth]{
  \begin{tabular}{l|c}
  %{0.61\width}
    \hline\hline
    %& \multicolumn{1}{c}{group~II disks} & \multicolumn{1}{c}{group~I disks}\\
    Parameter & Value  \\
    \hline
     $\gamma$       &  0.9   \\
     $\psi$         &  0.05  \\ 
     $h_{\rm c}$    &  0.075 \\ 
     $R_{\rm c}$     &  $125\,$au   \\ 
     $\Sigma_{\rm c}$
     $R_{\rm cav}$ & $0.41\,$au  \\
     %$M_{\mathrm{gas}} (\times10^{-2}$ M$_{\odot}$) & [1.34, 3.41, 6.8, 13.4, 34.1] & [4.7, 9.4, 23.5] \\
     %\mgas\,$\times10^{-2}\,$\msol & 1.4 - 70  \\
     \mgas & $6.7\times 10^{-2}\,$\msol  \\
     $M_{\rm dust}$ & $6.6\times 10^{-4}\,$\msol  \\
     \gdrat & $100$ \\
     $f_{\rm large}$ & 0.9 \\
     $\chi$ & 0.2 \\
     $L_{*}$ (L$_{\odot}$) & 37.7 \\
     $i$ ($^{\circ}$) & 45 \\
     d (pc) & 101 pc \\
    \hline
  \end{tabular}}
  \newline
    %\vspace{-0.05cm}
  %\begin{center}
  %\\
  %\footnotesize{\emph{Notes:} Values separated by a dash denote the parameter range that was explored.}
  %\end{center}
  %\caption{Parameters taken from \cite{deGregorioMonsalvoetal2013}. $^{\dagger}$In the models dust mass is kept fixed and the gas mass is varied, resulting in a varying $\Delta_{\rm gd}$. $^1$Spectrum from \citep{??}. }
\end{table}

HD\,163296 is one of the largest known disks, with a CO\,$J=3$\,--\,$2$ gas emission radius of $540\,$au \citep{Rosenfeldetal2013}.  Fitting of CO and $850\,\mu$m continuum emission, observed by ALMA, with a tapered surface density powerlaw yielded $\gamma = 0.9$ and $R_c = 125$ au \citep{Tillingetal2012, deGregorioMonsalvoetal2013}.  We model HD\,163296 with the stellar spectrum from the ProDiMo project \citep{Woitkeetal2019}, fixing the shape of the dust surface density profile to the above parameters and varying the gas mass.  To satisfy the radial profile of CO\,$3$\,--\,$2$ emission simultenaously with the spectral energy distribution, we find the density profile flaring index in Eq.~\ref{eq: gas density} is around $\psi=0.05$, consistent with the range of $0.019$ to $0.066$ found by \citet{Tillingetal2012}.  The morpology of the $^{12}$CO\,$3$\,--\,$2$ channel maps, in which both the near and far side of the disk can be seen, suggest HD\,163296 is more flared \citep[$\psi\approx0.12$,][]{ deGregorioMonsalvoetal2013} than our model ($\psi=0.05$).  However, these two $\psi$-s differ in physical meaning: the CO-based one measures the observed shape of the CO-emitting surface, while the disk structure parameter $\psi$ characterises the shape of the total gas mass distribution (see Eq.~\ref{eq: gas density}).

Our model which hits the HD upper limit reproduces the observed dust emission across the far-infrared and sub-millimetre wavelengths as well as various spatially resolved and unresolved emission lines of $^{12}$CO and its isotopologs, and has a gas-to-dust ratio \gdrat~$=100$.

Most previous estimates of the HD\,163296 gas mass relied on low-$J$ emission lines of CO isotopologs, and used a range of modelling approaches from generic model grids to tailored modelling with physical-chemical codes. Those \mgas\ estimates range from $8\times10^{-3}$ to $5.8\times 10^{-1}\,$\msol\ \citep{Isellaetal2007, WilliamsBest2014, Bonebergetal2016, Miotelloetal2016, WilliamsMcPartland2016, Powelletal2019, Woitkeetal2019, Boothetal2019}. The mass obtained from the most optically thin isotopolog among these, $^{13}$C$^{17}$O, was $2.1\times10^{-1}$\msol\ \citep{Boothetal2019}. 
%Using the errorbars on the isotopolog ratios assumed in that work\footnote{$^{12}$C$^{16}$O/$^{13}$C$^{16}$O~$=67\pm8$, $^{12}$C$^{16}$O/$^{12}$C$^{18}$O~$=444\pm88$, $^{12}$C$^{18}$O/$^{12}$C$^{17}$O~$=3.8\pm1.7$, see \citet{Boothetal2019}.}, we estimate a $1\sigma$ uncertainty of $50\,\%$ for their result, which makes it consistent within $2\sigma$ with our gas mass upper limit, $\mgas=6.7\times10^{-2}\,$\msol.
%The CO-based mass measurement closest to our upper limit is \mgas~$=4.8\times 10^{-2}\,$\msol, from spatially resolved $^{13}$CO and C$^{18}$O\,$2$--$1$ modelling by \citet{WilliamsMcPartland2016}.

Above, we assumed an undepleted solar abundance for elemental gas-phase carbon and oxygen.  Our model matching the HD upper limit over-produces the low-$J$ line fluxes of CO isotopologs by a factor of a few.  Since the rarer isotopologs are progressively more optically thin, we can reproduce their line fluxes by decreasing the gas-phase elemental carbon and oxygen abundance proportionately to the flux mismatch.  Since the millimetre-wave dust emission and HD upper limit constrain the gas-to-dust mass ratio to be $\leq 100$, we can combine the above considerations to arrive at three distinct hypotheses for HD\,163296:
\begin{enumerate}
    \item{\mgas\ is just sufficiently below our upper limit of $6.7\times10^{-2}\,$\msol\ for HD not to be detected.  If so, then as the dust mass is fixed, it follows from our models that \gdrat~$=100$ and that total gas-phase elemental C and O are depleted by up to a factor of a few.}
    \item{\mgas\ is a factor of a few below our HD limit, and the total gas-phase elemental C and O abundances are not depleted with respect to their interstellar values.  If so, the implication is that \gdrat~$\approx20$--$50$.  This relative depletion of gas over dust is supported by the hydrostatic MCMax modelling of the SED and low-$J$ $^{12}$CO, $^{13}$CO, and C$^{18}$O lines by \citet{Bonebergetal2016}, whose best models had $9.2<$~\gdrat~$<18$.  It is also consistent with the inner disk value \gdrat~$\approx55$, measured using accretion onto the central star by \citet[][their Figure~2]{Kamaetal2015}.}
    \item{\mgas\ is far below our upper limit.  In this hypothesis, the total C and O abundance in the gas must be enhanced above the interstellar baseline, in order to still match the optically thin CO isotopologs.  This would be the first known case of C and O enhancement, however the inner disk composition analysis by \citet{Kamaetal2015} does not show evidence for a strong enhancement of gas-phase volatile elements over total hydrogen.}
\end{enumerate}

%The combined evidence thus favours the second hypothesis: that \mgas~$\approx (1-3)\times10^{-2}$ and thus \gdrat~$\approx(20-50)$.
Thus \gdrat~$>100$ is ruled out by the HD\,$1$\,--\,$0$ upper limit for HD 163296, independently of assumptions about the precise abundance of gas-phase volatiles.

The abundance of volatile elements in the HD\,163296 disk may be depleted or enhanced by up to a factor of a few, depending on the true value of \mgas\ and on the somewhat uncertain underlying number abundance ratios of $^{12}$CO and its various isotopologs.  We note that even with the flat, cold disk structure of HD\,163296, our \gdrat~$=100$ model somewhat over-predicts the CO emission outside of $\sim100\,$au for an undepleted elemental carbon abundance (C/H~$=1.35\times 10^{-4}$).  A more flared surface would aggravate this over-prediction, while globally reducing the elemental C under-predicts the CO\,$3$\,--\,$2$ inside $\sim100\,$au.  This may indicate that any depletion of gas-phase volatile elemental C and O, reflected in the CO abundance in the warm molecular layer, is restricted to the region beyond the CO snowline, which has been observed to be at $\approx90\,$au \citep{Qietal2015}.  The same conclusion was recently reached by \citet{Zhangetal2019} through an analysis of spatially resolved C$^{18}$O data, which yielded a factor of ten depletion of gas-phase CO outside the CO snowline.

\subsection{HD\,100546}

HD\,100546 was previously modelled with DALI by \citet{Brudereretal2012} who determined the radial and vertical structure of the disk mainly from CO lines and continuum emission.  A refined version of this modelling effort included the \emph{Herschel} HD upper limits, the C$^{0}$ and C$_{2}$H fluxes, and the spatially resolved CO\,$3$\,--\,$2$ emission, constraining the gas mass to $8.1\times10^{-3}\leq$~\mgas~$\leq2.4\times10^{-1}\,$\msol\ \citep{Kamaetal2016b}.  The highest-mass model had \gdrat~$=300$, with a dust mass anchored by the continuum spectral energy distribution.  Due to a factor of four error in the D abundance used in that model, we revise those numbers to $\lesssim100$ and thus \mgas~$\lesssim0.08\,$\msol\ from the \citet{Kamaetal2016b} model.  This is about a factor of two stronger than the constraint from our general model grid, so in Figure~\ref{fig:mass_summary} we adopt \mgas~$\lesssim0.08\,$\msol. 

%Our gas mass constraint for HD\,100546 in Figure~\ref{fig:hdfluxes 1-0} is less strict than that in \citet{Kamaetal2016b}
%A detailed modelling continuum and line data, including the HD\,$1$--$0$ upper limit, for HD\,100546 previously constrained its gas-to-dust mass ratio to $\Delta_{\rm gd} = 10-300$ \citep{Kamaetal2016b}.

\subsection{Other individual disks}

%HD 144668 has weak HD gas mass constraints and no relevant dust observation/ dust mass

\hspace{\parindent}{\bf HD\,97048} hosts a massive dust disk, $\mdust\simeq6.7 \times 10^{-4}\,$\msol\ \citep{Walshetal2016}, so it is likely the gas mass is also high. The disk surface is highly flared \citep[$\Psi = 0.5\,-\,0.73$, see e.g.][]{Lagageetal2006, Walshetal2016, Ginskietal2016,vanderPlasetal2019}). This exceeds the largest $\Psi$ in our general grid, but we note again that the CO-surface $\Psi$ and the density structure $\Psi$ differ in physical meaning.
From our grid we find $\mgas \leq9.4\cdot10^{-2}\,$\msol\ ($\gdrat \leq 200$). 
%%% Note from Mihkel: I left out the very high Psi results. It's reasonable that the mass constraint for a warmer disk is stronger, but these super-high psi's aren't really physically meaningful for a hydrostatic disk structure so I don't think we want to open this can of worms in this paper.
%Tests show that if the disk more flared ($\Psi = 0.5-0.73$), the upper limit on the gas mass is lower: $\mgas \leq 4.7\cdot10^{-2}\,$\msol\  (\gdrat~$\leq100)$.  However, we note that the physical meaning of the flaring parameter $\psi$ describing the PAH emission, CO-emitting surface or scattered light surface is different from the $\psi$ describing the gas density structure.

\textbf{HD\,104237. }  For this disk, \citet{Halesetal2014} determined  $\mdust=4\times10^{-4}\,$\msol, which assuming $\gdrat=100$ implies a total mass $\mgas=4\times10^{-2}\,$\msol. This is consistent with our upper limit from HD\,$1$\,--\,$0$, which yields an upper limit of $\gdrat \leq 300$ (Figure~\ref{fig:HD10_Mgas_1mm}). 

\textbf{HD\,36112 (MWC 758). }Based on millimetre continuum interferometry, \citet{Guilloteauetal2011} inferred a disk mass of $(1.1\pm0.2)\times 10^{-2}\,$\msol. Our analysis of the $1.3\,$mm continuum flux and the HD\,$1$\,--\,$0$ upper limit matches both datapoints for \gdrat~$\approx 100$ and a disk mass of order $10^{-2}\,$\msol. A substantially lower gas mass would imply a very low \gdrat\ mass ratio.

\textbf{HD\,31293 (AB Aurigae). } From 1.3 millimetre continuum observations performed using the SMA, \citet{Andrewsetal2013} inferred a dust mass of $(1.56\pm0.09)\times 10^{-4}\,$\msol, implying $M_{\rm gas} = 1.56\times 10^{-2}\,$\msol\ assuming \gdrat = 100. The high upper limit of HD 1\,-\,0 for this source does not allow us to put any meaningful constraints on the gas mass based on HD.

\textbf{HD\,135344B} has been modelled by \citet{vdMarel2016} to determine the physical structure. Using ALMA observations of $^{13}$CO $J$\,=\,3\,--\,2, C$^{18}$O $J$\,=\,3\,--\,2, $^{12}$CO $J$\,=\,6\,--\,5 and dust 690 GHz continuum, they determined a gas mass $M_{\rm gas} = 1.5\times10^{-2}\ \mathrm{M}_{\odot}$. We run models based on their physical structure and find the resulting HD\,$1$\,--\,$0$ flux to be in agreement with the upper limit (see Figure \ref{fig: HD135344B fit} in Appendix \ref{app: HD135344B}).
%HD 139614 has a lot of IR VLTI observations showing a 6 AU gap, but no millimetre dust/line observations published. The disk is covered in Olja's project, so observations will be forthcoming

\textbf{HD\,142527. } Modelling interferometric $880\,\mu$m continuum and $^{13}$CO\,$3$--$2$ and C$^{18}$O\,$3$--$2$ line observations, \citet{Boehleretal2017} determine a dust mass of $1.5\times10^{-3}\,$\msol\ and a gas mass of $5.7\times10^{-3}\,$\msol\ \citep[see also][]{Mutoetal2015}.  This gives $3\leq$~\gdrat~$\leq5$ and suggests the gas is either strongly depleted in elemental C and O, or dissipating entirely.  Due to the loose HD\,$1$\,--\,$0$ upper limit for this source, we cannot provide an independent check of the low \gdrat\ derived from CO.
%\lt{-- Ewine mentioned that we should check the gas mass derivation using Anna's models, but I suggest we keep that for after submitting --}

%HD 179218 millimetre observations are from 2000 (Mannings&Sargent), so we only have a luminosity based mass estimate. No ALMA observations are forthcoming
\textbf{HD\,179218. }From the integrated 1.3 millimetre flux \citet{ManningsSargent2000} infer a dust mass of ${(1.5\pm0.15)\times 10^{-4}\,}$\msol, implying $M_{\rm gas} = 1.5\times 10^{-2}\,$\msol\ assuming \gdrat = 100. Again the HD\,$1$\,--\,$0$ upper limit provides no meaningful constraint on the gas mass.

\textbf{HD\,100453. }Based on millimetre continuum interfermotric observations, \citet{vanderPlasetal2019} inferred a dust mass of $6.7\times10^{-5}\,$\msol. By comparing the $^{13}$CO 2\,--\,1 and C$^{18}$O 2\,--\,1 to the disk model grid in \cite{WilliamsBest2014}, they determine a gas mass of $(1-3\times)\times 10^{-3}\,$\msol. Combining both disk masses implies a gas-to-dust mass ratio of \gdrat~$15-45$. From our analysis of the $1.3\,$mm continuum flux and the HD\,$1$\,--\,$0$ upper limit we constrain gas mass to $M_{\rm gas}\leq 10^{-2}\,$\msol\ and the gas-to-dust mass ratio \gdrat~$\leq300$. Both constraints are in agreement with the results of \cite{vanderPlasetal2019}.

\textbf{HD\,169142. }From interferometric 1.3 millimetre continuum and $^{12}$CO 2\,--\,1, $^{13}$CO 2\,--\,1 and C$^{18}$O 2\,--\,1 line observations, \citet{Panicetal2008} derived a dust mass of $2.16\times10^{-4}\,$\msol\ and a gas mass of $(0.6-3.0)\times10^{-2}\,$\msol. \cite{Fedeleetal2017} find similar disk masses based on higher resolution observations. Constraints based on our analysis of the $1.3\,$mm continuum flux and the HD\,$1$\,--\,$0$ upper limit put the gas mass at $M_{\rm gas} \leq 4\times10^{-2}\,$\msol\ and \gdrat~$\leq300$, both of which are in good agreement with previous results.

\textbf{Oph IRS 48 (WLY 2-48). }\cite{vdMarel2016} modelled the resolved $440\,\mu$m continuum and $^{13}$CO 6\,--\,5 and C$^{18}$O 6\,--\,5 line observations. They derived a dust mass of $1.5\times10^{-5}\,$\msol\ and a gas mass of $5.5\times10^{-4}\,$\msol, giving a gas-to-dust mass ratio of \gdrat~$\approx 37$. Constraints from our analysis of the HD\,$1$\,--\,$0$ line flux and 1.3 millimetre continuum give $\mgas \lessapprox 10^{-2}\,$\msol\ and \gdrat~$\lessapprox 300$. These upper limits agree with previous results.

\subsection{Are the disks gravitationally stable?}
\label{sec:stability}

Constraints on \mgas\ allow to test whether the disks in our sample are currently gravitationally stable. Gravitational instability, leading to spirals or fragmentation, occurs in disk regions which are dense and cold, and have low orbital shearing on the timescale of the instability (i.e. at large radii). This is quantified with the Toomre $Q$ parameter, $Q = \Omega_{K}\,c_{\rm s}\, (\pi\,G\,\Sigma)^{-1}$ \citep{Toomre1964}, which simplifies to
\begin{equation}
\label{eq: instability}
Q = 21\times \left(\frac{\Sigma}{10\,{\rm kg\,m^{-2}}}\right)^{-1}\,\times \left(\frac{r}{100\,{\rm au}}\right)^{-3/2},
\end{equation}
following \citet{KimuraTsuribe2012}. If $Q<1$, the disk will fragment. For $1<Q<2$, the disk will be marginally stable, developing transient spirals and clumps, while for $Q>2$ it is stable against gravitational collapse. Assuming a surface density profile $\Sigma=\Sigma_{0}\times(r/r_{0})^{-1}$ and $M_{\rm disk}\approx M_{\rm gas}$, we obtain
\begin{equation}
Q = 2.44\times10^{22}\,\pi\,r_{0}^{1/2}\,M_{\rm disk}^{-1},
\end{equation}

Our most massive disk models have \mgas~$=0.1\,$\msol. Taking a characteristic radius $r_{0}=100\,$au, we find $Q=1.5$, which is marginally stable. The disks for which we have the weakest upper limits relative to the massive disk models -- HD\,142527, HD\,144668, HD\,179218, and HD\,31293 -- may potentially be gravitationally unstable within the limits of the \emph{Herschel} HD data.  For the rest of the sample, a gravitationally unstable \mgas\ is effectively ruled out, i.e. they are most likely stable.

Dust dips, rings, or cavities may locally affect the temperature structure of the gas and thus, through the sound speed, the local $Q$ in a disk ($Q\propto \tkin^{0.5}$, i.e. a weak dependence). In general, a lower dust surface density leads to more efficient external heating and thus more stability. Inside a local dust enhancement, the temperature may drop somewhat, but if the region is already very optically thick, the effect on \tkin\ will be minor. We therefore have not considered the effect of such $\Sigma_{\rm dust}$ perturbations in this paper.

\begin{figure*}[!ht]
\centering
\includegraphics[clip=,width=1.0\linewidth]{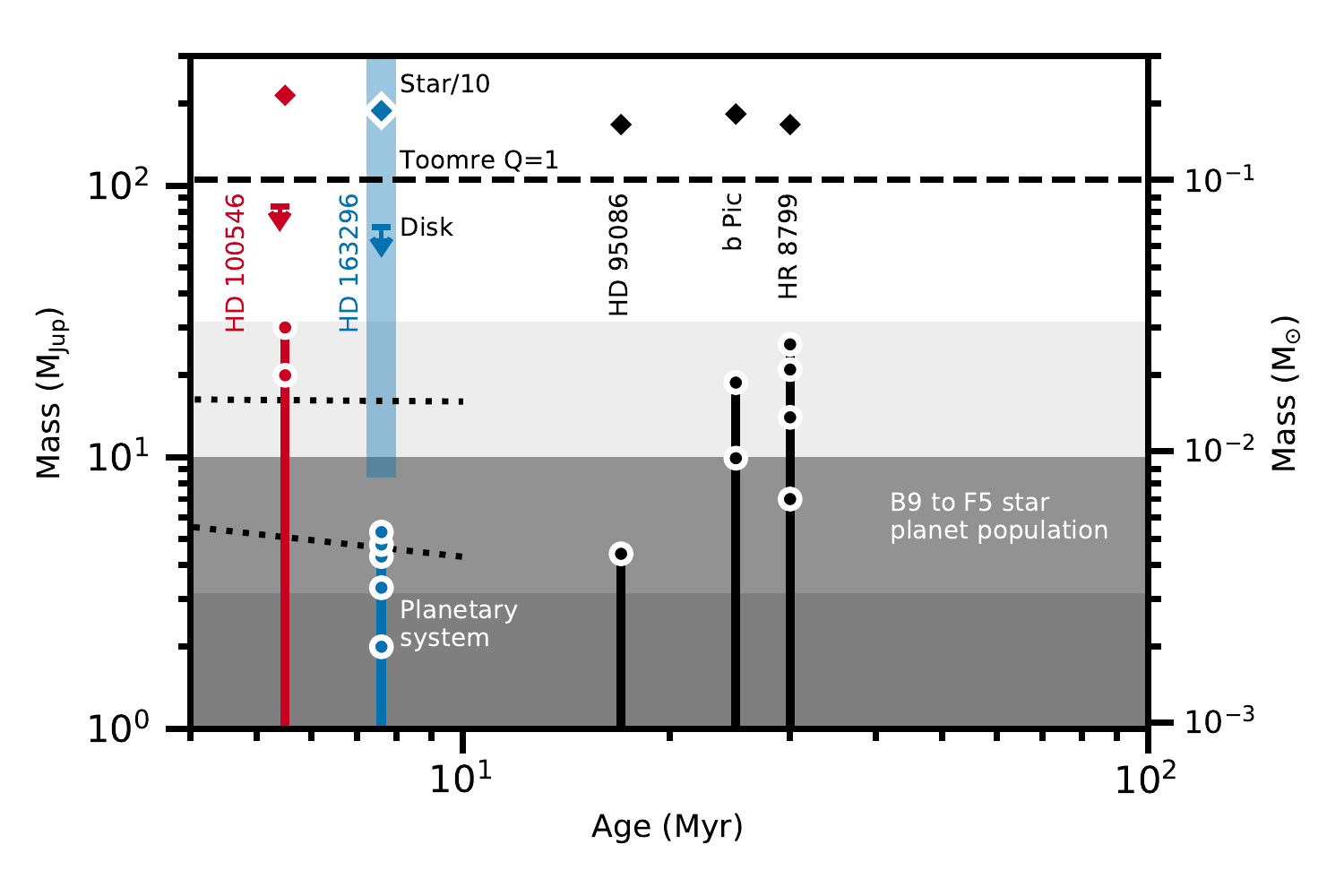}
\caption{Mass of selected disks and planets around B9 to F5 type stars.  Vertical lines show the cumulative mass of each planetary system, with dots highlighting planets from the most massive at bottom.  Disk gas mass upper limits from HD lines are from this work (HD\,163296) and from \citet[][HD\,100546]{Kamaetal2016b}.  For HD\,163296, the range of CO-isotopolog based disk mass estimates is shown by a light blue bar ($8\times10^{-3}$ to $5.8\times10^{-1}\,$\msol; references in text).  Also shown are the stellar mass divided by 10 and age; the mass limit for a gravitationally unstable disk (dashed line); an extrapolated dust-based disk mass range \citep[dotted lines,][]{Pascuccietal2016}; and a population density colormap for planets around B9 to F5 type stars (data retrieved from \texttt{exoplanets.org} on 2019.07.16; bins contain from bottom to top 7, 6, and 1 planet).  See text for individual planet and stellar mass references.}
%Disk mass upper limits \citep[red and blue arrows for flared and flat disks; this work and][]{Kamaetal2016b}, stellar masses divided by $10$ \citep[diamonds,][and tbd]{Vioqueetal2018}, and the total mass of known planets in each system \citep[filled circles,][and tbd]{Teagueetal2018,Pinteetal2018}.  }
% Masses of individual planetary systems are compiled from \citet{Maroisetal2010}, \citet{Pinillaetal2015}, \citet{DeRosaetal2016}, \citet{Teagueetal2018}, \citet{Pinteetal2018}, and \citet{Lagrangeetal2019}.
\label{fig:mass_summary}
\end{figure*}

\section{Discussion}\label{sec:discussion}
%\textbf{|MK: review survey results of CO-based gas masses in intro; also review importance of high gas mass in general (Toomre Q unstable; also total gas reservoir for giant planets; elemental abundance reference point)|}

\subsection{Mass of disks, stars, and planets}

Intermediate-mass stars (spectral types B9 to F5, masses $1.5$ to $3\,$\msol) host some of the best-studied protoplanetary disks and high-mass planetary systems.  Several Herbig~Ae/Be protoplanetary disks have also yielded detections of protoplanet candidates.  This presents an opportunity to investigate equivalent planetary systems at different stages of evolution.

In Figure~\ref{fig:mass_summary}, we compare the disk gas mass with the host star and the candidate protoplanets in the disk. We show two Herbig~Ae/Be systems with strong mass limits, HD\,163296 ($\mgas\leq0.067\,$\msol, this work) and HD\,100546 \citep[$\mgas\leq0.08\,$\msol,][]{Kamaetal2016b}. Much of the work on embedded protoplanet candidates quoted below is very new. There are large, at least a factor of two to ten, uncertainties behind the planet mass estimates below, in particular for those inferred from dust gaps where the $\alpha$ viscosity parameter plays a role. We adopt middle-ground values from the literature to begin a discussion on comparing disk and embedded planet masses.

For HD\,163296, our HD-based upper limit rules out a large fraction of the wide range of CO isotopolog based \mgas\ estimates from the literature. Of those still possible, the lowest is $\mgas=8\times10^{-3}\,$\msol. The presence of five giant planets has been inferred from dust gaps and gas kinematics: at $10\,$au with a mass $(0.53\pm0.18)\,$\mjup\ for $\alpha_{\rm visc}=10^{-4}$ to $10^{-3}$ \citep{Zhangetal2018}; at $48\,$au with $0.46\,$\mjup\ \citep{Isellaetal2016, Liuetal2018}; at $86\,$au with $(1\pm0.5)\,$\mjup\ \citep{Liuetal2018, Teagueetal2018}; at $145\,$au with $1.3\,$\mjup\ \citep{Liuetal2018, Teagueetal2018}; and at $260\,$au with $2\,$\mjup\ \citep{Pinteetal2018}. Using the HD- and CO-based \mgas\ limits, and taking the combined mass of all published protoplanets in this disk as $\approx 5\,$\mjup, we find the HD\,163296 disk has converted $10$ to $40\,$\% of its mass into giant planets.

%\footnote{We refer here to the present-day mass. For the $40\,$\% mass conversion, we add the CO-based \mgas\ and the total mass of the planets ($\approx 5\,$\mjup).} 

For HD\,100546, the planet masses were constrained to be $\approx 10\,$\mjup\ at $10\,$au and $\sim10\,$\mjup\ at $70\,$au by \citet{Pinillaetal2015}. The mass of the outer planet could be $<5\,$\mjup\ ($>15\,$\mjup) if it formed very early (late), so we adopt $10\,$\mjup. The HD-based \mgas\ upper limit and the combined mass of the candidate planets yield a lower limit on the disk-to-planet mass conversion efficiency, $\gtrsim 30\,$\%.

Such high disk-to-planet mass conversion efficiencies combined with the presence of several gas giants per star raise the question of whether the planets formed through gravitational instability.  Adding the \mgas\ upper limit and combined mass of proposed planets in either disk gives a result close to $0.1\,$\msol.  This is approximately at the gravitationally unstable limit, so such a formation pathway may be feasible even with the current total mass in the system, although the local Toomre $Q$ varies with radius and may leave the outer disk still far from instability \citep[e.g.][]{Boothetal2019}.
%if the true disk mass in either case is not substantially below the HD-based upper limit. Furthermore, in their early evolution the disks were likely more massive.
%Otherwise, core accretion models will have to explain the formation of $\sim1\,$\mjup\ objects at orbital distances up to $260\,$au within $5\,$Myr.

We also show in Figure~\ref{fig:mass_summary} three somewhat older stars of similar mass (HD\,95086, $\beta\,$Pic, and HR\,8799) and their planets; standard disk mass estimates for stars of $1.5$ and $3\,$\msol\ based on \mdust\ relations from \citet{Pascuccietal2016} and scaled up with $\gdrat=100$; and a shaded log-scale histogram of the mass distribution of known planets around early-type stars\footnote{Planets retrieved from \texttt{exoplanets.org} on 2019.07.16.}.  Stellar masses are from the GAIA DR2 analysis by \citet{Vioqueetal2018}, and from \citet[][$\beta\,$Pic]{Davidetal2015} and \citet[][HD\,95086]{Stassunetal2018}.  Planet masses for individual systems are plotted as cumulative bars, with the highest-mass planet at the base. We compiled planet data from \citet{Teagueetal2018}, \citet{Pinteetal2018,Pinteetal2019}, \citet{Pinillaetal2015}, \citet{Liuetal2018}, \citet{Zhangetal2018}, \citet{Rameauetal2013b,Rameauetal2013c}, \citet{DeRosaetal2016}, and \citet{Maroisetal2008,Maroisetal2010}.  Individual stellar masses are from \citet{Rheeetal2007}, \citet{Davidetal2015}, \citet{Stassunetal2018}, and \citet{Vioqueetal2018}.
%A comparison of disk mass limits and planetary system masses will provide a measure of planet formation efficiency from a mass-conversion point of view.
%\mk{Left off here 2019.10.25 15:47, will pick up later tonight :)}

The two HD-based disk \mgas\ limits in Fig.~\ref{fig:mass_summary} exceed the combined mass of planets around HR\,8799, the most massive known planetary system, by a factor of only three.  The disk mass limits are also only a factor of three above combined mass of candidate protoplanets in the HD\,100546 disk.  Either A-type star disks can, in some cases, convert $10\,$\% or more of their mass into giant planets, or these planetary systems formed at a very early stage, perhaps while the central protostar and massive initial disk were still heavily accreting from the protostellar envelope in which they were embedded. The mass distribution of giant planets around main-sequence A and B stars (Fig.~\ref{fig:mass_summary}) is strongly skewed towards lower masses, suggesting that such extreme mass conversion events are either rare, or that the high-mass planetary systems are not stable on timescales beyond a few times $10\,$Myr.

\begin{figure*}[!th]
\centering
\includegraphics[clip=,width=1.0\linewidth]{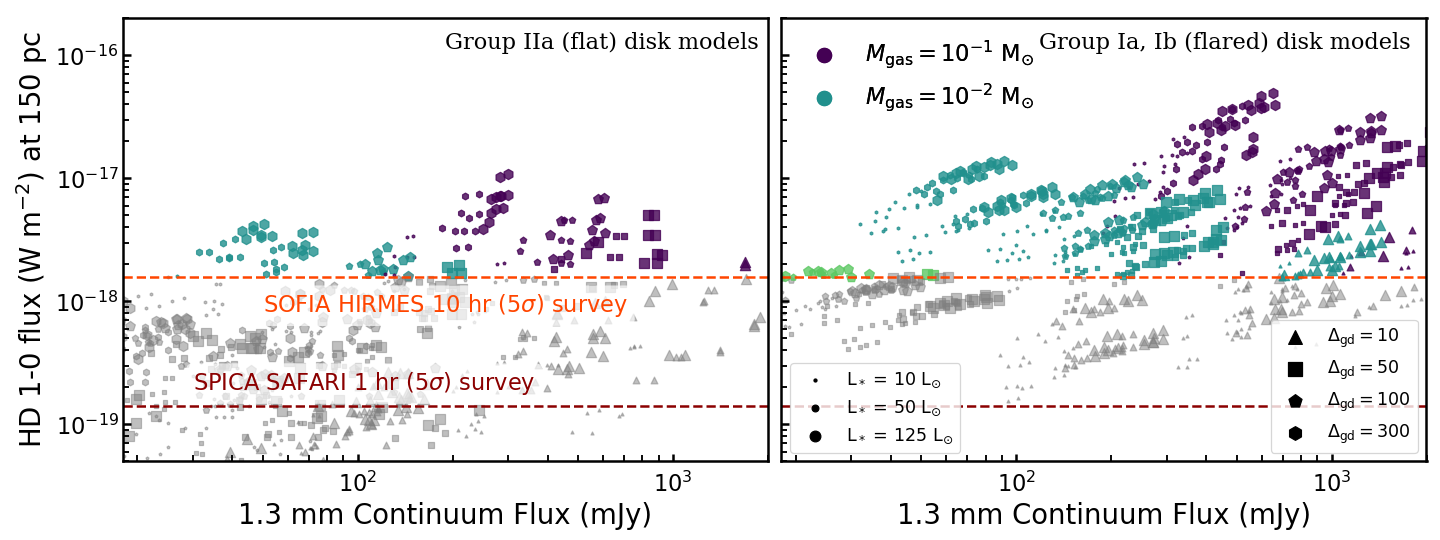}
\caption{\label{fig: sofia hirmes} Observability of Group Ia,Ib (left) and Group IIa (right) models with SOFIA HIRMES. Coloured disk models are detectable $(\geq5\sigma)$ with a 10 hr integration. Dark red dashed line shows the SPICA/SAFARI 1 hr detection limit. The \emph{Origins Space Telescope} 1 hr detection limit ($\sim1\times10^{-20}$ W m$^{-2}$) lies below the limits of the figure.
Note that the fluxes are calculated for a distance of 150 pc.}
\end{figure*}

\subsection{Observing HD in Herbig disks with SOFIA/HIRMES, SPICA/SAFARI and emph{Origins Space Telescope}}
\label{sec: sofia and spica}

In the coming years, several facilities will or may become available for observing HD rotational lines. The HIRMES instrument for SOFIA is currently undergoing commissioning and is due to be delivered at the end of 2020 \citep{Richardsetal2018}.
HIRMES will have a high spectral resolution of R~$\sim100000$, allowing us, for the first time to spectrally resolve the HD 1\,-\,0 line. The sensitivity of HIRMES will be similar to Herschel/PACS.
Our models suggest some Herbig~Ae/Be disks will be detectable with this instrument, assuming the necessary hours per source are available.

Figure~\ref{fig: sofia hirmes} shows the detectability of our disk models with a $10\,$h SOFIA/HIRMES observation, assuming a distance of $150\,$pc. Of the flat models (group~II disks), only the most massive (\mgas~$\sim0.1\,$\msol) around stars with the highest stellar luminosity ($L_* \geq 50\,$\lsol) are detectable. Among the flared models (group~I), a larger fraction of disks is observable. All of the disk models $M_{\rm gas} = 0.1\ \mathrm{M}_{\odot}$ where $\Delta_{\rm gd} > 10$ should be detectable in 10 hrs with SOFIA/HIRMES. For those disks with $M_{\rm gas} = 0.01\ \mathrm{M}_{\odot}$, all systems with $L_* = 125\ \mathrm{L}_{\odot}$ and most systems with $L_* = 50\ \mathrm{L}_{\odot}$ are detectable. To maximize the chance of success, future SOFIA/HIRMES observations should select group~I sources with high stellar luminosity.

Based on the stellar luminosities in Table \ref{tab:obs} there are four group I sources that match these criteria best for SOFIA/HIRMES to detect the HD\,$1$\,--\,$0$ line: HD\,31293 (AB Aur), HD\,100546, HD\,179218 and HD\,97048. 
For these sources a $10\,$h observation with SOFIA/HIRMES would improve the current upper limits by a factor $3$--$10$ and constrain the gas-to-dust mass ratio to \gdrat~$\leq 50-100$ if the sources remain undetected. 

Beyond SOFIA/HIRMES there are two proposed space missions focussing on far-infrared observations: SPICA/SAFARI and \emph{Origin Space Telescope}.  SPICA is one of the competitors for ESA's M5 opportunity, with a resolving power R~$\sim3000$ and a $5\sigma$ 1 hr sensitivity of $1.3\times10^{-19}\,$W$\,$m$^{-2}$ at $112\,\mu$m \citep{Audleyetal2018} . The \emph{Origin Space Telescope} is a NASA mission concept. It would have high spectral resolution (R~$\sim43000$) and sensitivity ($\sim1\times10^{-20}\,$W$\,$m$^{-2}$ in 1 hr) at $112\,\mu$m \citep{Bonatoetal2019}.  Hydrogen deuteride in all Herbig~Ae/Be disks, and many T~Tauris, within $\sim200\,$pc will be detectable with these missions.  However, both still require final approval and would only become available at the end of the 2020's at the earliest. If approved, these missions would be an enormous step forward in planet-forming disk studies.

\section{Conclusions}
\label{sec: conclusion}

%Putting constraints on the total gas mass of protoplanetary disks is crucial for improving our understanding of disk evolution and planet formation. The most precise gas disk mass measurements come from HD rotational lines (e.g., \citealt{Berginetal2013, McClureetal2016,Trapmanetal2017}), but detections of HD are so far limited to three T-Tauri disks. The DIGIT key programme of \emph{Herschel} covered the HD $J$\,=\,1\,-\,0 and $J$\,=\,2\,-\,1 rotational lines for a number of disks around Herbig~Ae/Be stars, resulting in no detections. In this work, we combine the observed upper limits of the HD 1\,-\,0 line with an extensive grid of herbig disk models run using the thermochemical code \texttt{DALI} \citep{Brudereretal2012,Brudereretal2013} to put upper limits on the gas masses of 15 Herbig~Ae/Be disks. Our main conclusions are:
\begin{enumerate}
%    \item For the flared, group I sources in our sample we find \mgas~$<0.02$--$0.03\,$\msol\ for IRS\,48, HD\,36112, HD\,100453, and HD\,135344B. For HD\,139614 and HD\,100546, the upper limit is somewhat higher, \mgas~$\lesssim 0.2$--$0.3\,$\msol. Constraints for the flat group II sources are weaker due to the overall lower brightness of these cold disks.
    \item{We find an overall gas mass upper limit of \mgas~$\leq0.1\,$\msol\ for most of the disks studied.  None of the disks are very likely to be strongly gravitationally unstable, although the constraints for HD\,142527, HD\,144668, HD\,179218, and HD\,31293 (AB\,Aur) are weak enough to allow for the possibility.}
    
    \item{The HD\,163296 disk mass is $\mgas\leq6.7\times10^{-2}\,$\msol, based on the HD\,$1$\,--\,$0$ upper limit. The CO-based literature lower limit is $\mgas=8\times10^{-3}\,$\msol, contingent on the true level of gas-phase volatile depletion. The gas-to-dust ratio is thus $12\leq\gdrat\leq100$, indicating gas dissipation may be proceeding faster than dust removal in this disk. This is consistent with $\gdrat=55$ inferred from the accretion-contaminated photosphere of the central star \citep{Kamaetal2015}.}
    %\item For HD\,97048 we combine the gas mass upper limit from HD 1\,-\,0 with the lower limit inferred from CO isotopologues to constrain it gas mass to  $M_{\rm gas} \leq 0.05-0.1\ \mathrm{M}_{\odot}$.
    
    \item{Comparing the HD\,163296 and HD\,100546 \mgas\ constraints with their protoplanet candidates and the HR\,8799 giant planet system, we find that at least some Herbig~Ae/Be disks convert the equivalent of $10$ to $40\,$\% of their present-day mass into giant planets.}
    
    %\item{\lt{This conclusion is no longer a (major) part of the paper} The HD\,97048 disk mass is \mgas~$\leq0.1\,$\msol\. However, if the disk is extremely flared ($\psi\approx0.3$), as scattered light and CO rotational line imaging suggest, the upper limit is lowered to \mgas~$\geq0.05\,$\msol.}
    
    \item{Near-future SOFIA/HIRMES observations will probe the mass of flaring disks and large flat disks around A-type stars within $\approx150\,$pc with $\gtrsim10\,$h integrations.  SPICA/SAFARI will be crucial for larger sample studies of \mgas\ in disks. OST, if approved, would further revolutionise the field.} 
\end{enumerate}

\begin{acknowledgements}
We thank Sebastiaan Krijt for useful discussions. 
MK gratefully acknowledges funding from the European Union's Horizon 2020 research and innovation programme under the Marie Sklodowska-Curie Fellowship grant agreement No 753799. LT is supported by NWO grant 614.001.352. DF acknowledges financial support provided by the Italian Ministry of Education, Universities and Research, project SIR (RBSI14ZRHR). AM and CC gratefully acknowledge funding from the European Union’s Horizon 2020 research and innovation programme under the Marie Sklodowska-Curie grant agreement No
823823, (RISE DUSTBUSTERS), and AM funding by the Deutsche Forschungsgemeinschaft (DFG, German Research Foundation) – Ref no. FOR2634/1 ER685/11-1. EAB gratefully acknowledges support from NASA via grant NNX16AB48G-XRP. All figures were generated with the \texttt{PYTHON}-based package \texttt{MATPLOTLIB} \citep{Hunter2007}.
\end{acknowledgements}

\bibliographystyle{aa}
\bibliography{gasmass}

\begin{appendix}

\newpage
\section{HD 1\,-\,0 fluxes for HD\,135344B}
\label{app: HD135344B}

\begin{figure}[ht]
\centering
\includegraphics[clip=,width=1.0\columnwidth]{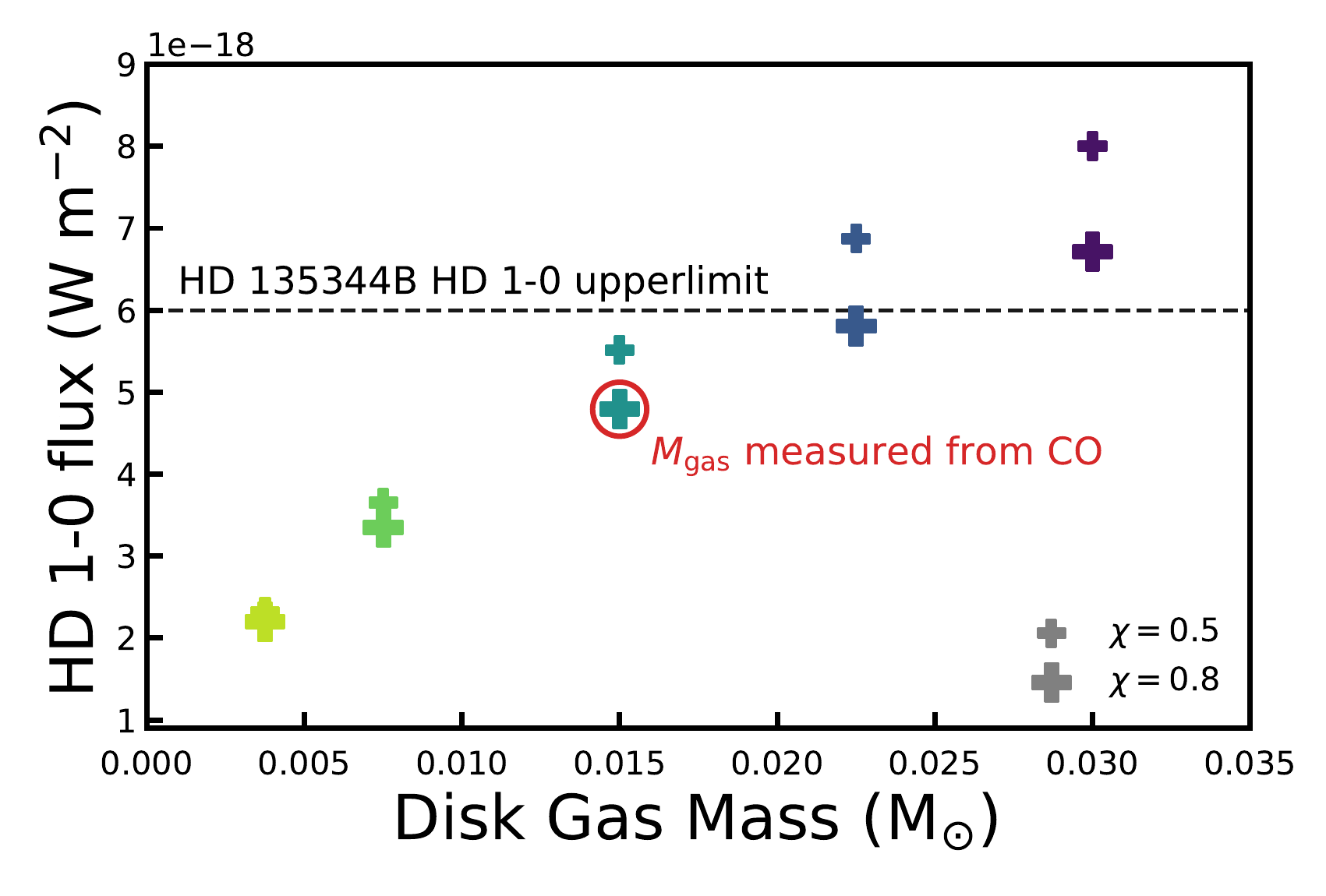}
\caption{\label{fig: HD135344B fit} Comparing the HD135344B specific models from \citet{vdMarel2016} to HD 1\,-\,0 upper limit \citep{Fedeleetal2013}. All models have a dust mass $M_{\rm dust} = 1.3\times10^{-4}\ \mathrm{M}_{\odot}$ (cf. Table 3 in \citealt{vdMarel2016}). The red circle shows the gas mass inferred from CO by \cite{vdMarel2016}. }
\end{figure}

Based on the HD135344B source-specific model from \cite{vdMarel2015,vdMarel2016}, we run a series of 10 models, varying the disk gas mass between $3.75\times10^{-3}\ \mathrm{M}_{\odot}$ and $3\times10^{-2}\ \mathrm{M}_{\odot}$. Figure \ref{fig: HD135344B fit} compares the HD 1\,-\,0 line fluxes of these models to the observed upper limit (Table \ref{tab:obs}). From the CO isotopolog observations \cite{vdMarel2016} infer $M_{\rm gas} = 1.5\times10^{-2}\ \mathrm{M}_{\odot}$. This gas mass is in agreement with the gas mass upper limit inferred from HD 1\,-\,0, $M_{\rm gas} \leq 2.3\times10^{-2}\ \mathrm{M}_{\odot}$. Note that both gas masses are much lower than $0.1\ \mathrm{M}_{\odot}$, making it highly unlikely that HD\,135344B is gravitationally unstable (Section~\ref{sec:stability}).

\section{HD 2\,-\,1 upper limits versus the model fluxes}
\label{app: HD 2-1 upper limits}

\begin{figure*}[!h]
\centering
\begin{subfigure}{0.99\linewidth}
\centering
\includegraphics[clip=,width=1.0\linewidth]{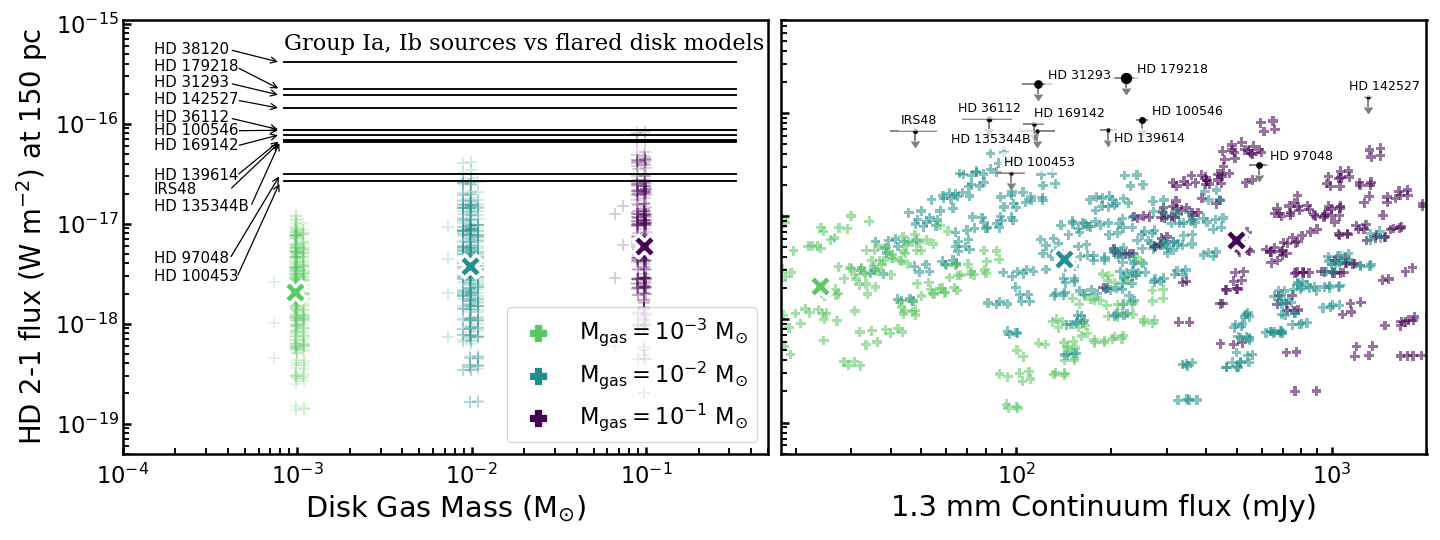}
\end{subfigure}
\begin{subfigure}{0.99\linewidth}
\centering
\includegraphics[clip=,width=1.0\linewidth]{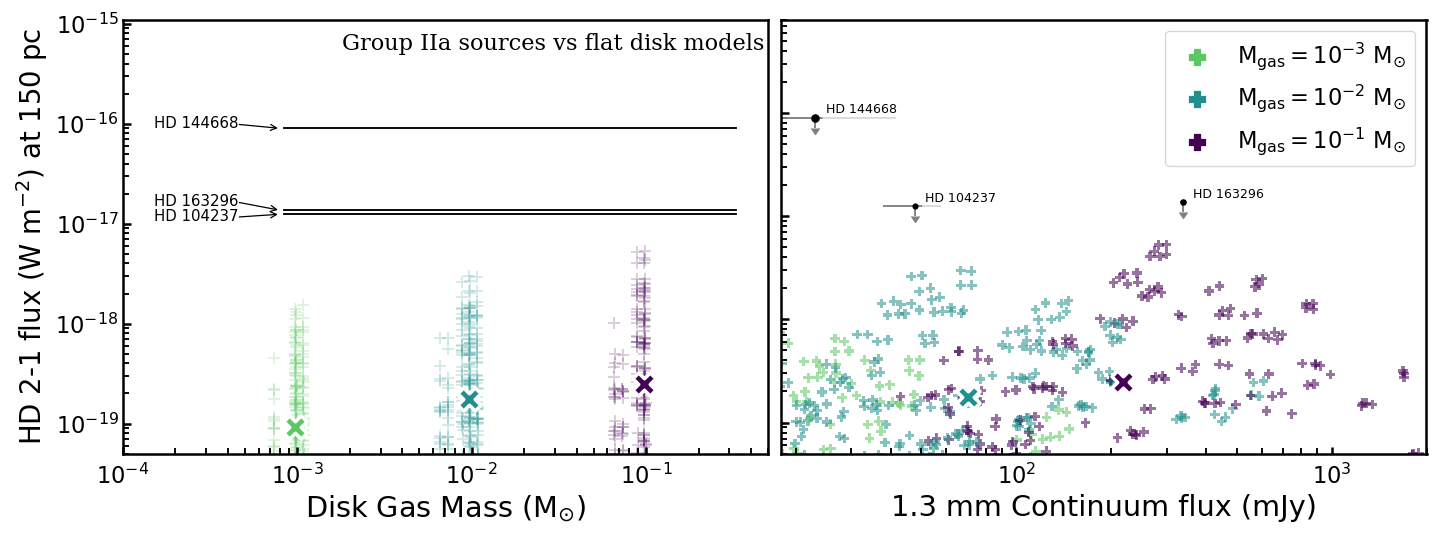}
\end{subfigure}
\caption{Upper limits on HD~$56\,\mu$m line flux for the sample of Herbig Ae/Be disk systems (black lines) compared with our grid of DALI disk models (crosses). The top panels show the group I sources compared to models with flaring angle $\psi = 0.3$. The bottom panels show the group II sources compared to models with $\psi = 0.0$. \textbf{Left:} models are separated based on gas mass. 
\textbf{Right:} HD 2\,-\,1 upper limits set against 1.3 mm continuum fluxes for both observations and models.  }
\label{fig:hdfluxes 2-1}
\end{figure*}

\section{HD 1\,-\,0 line versus 1.3 mm continuum fluxes, showing gas-to-dust ratios and stellar luminosities}
\label{app: HD 1-0 vs 1.3 mm -- extended}

\begin{figure}[!ht]
\centering
\includegraphics[clip=,width=1.0\columnwidth]{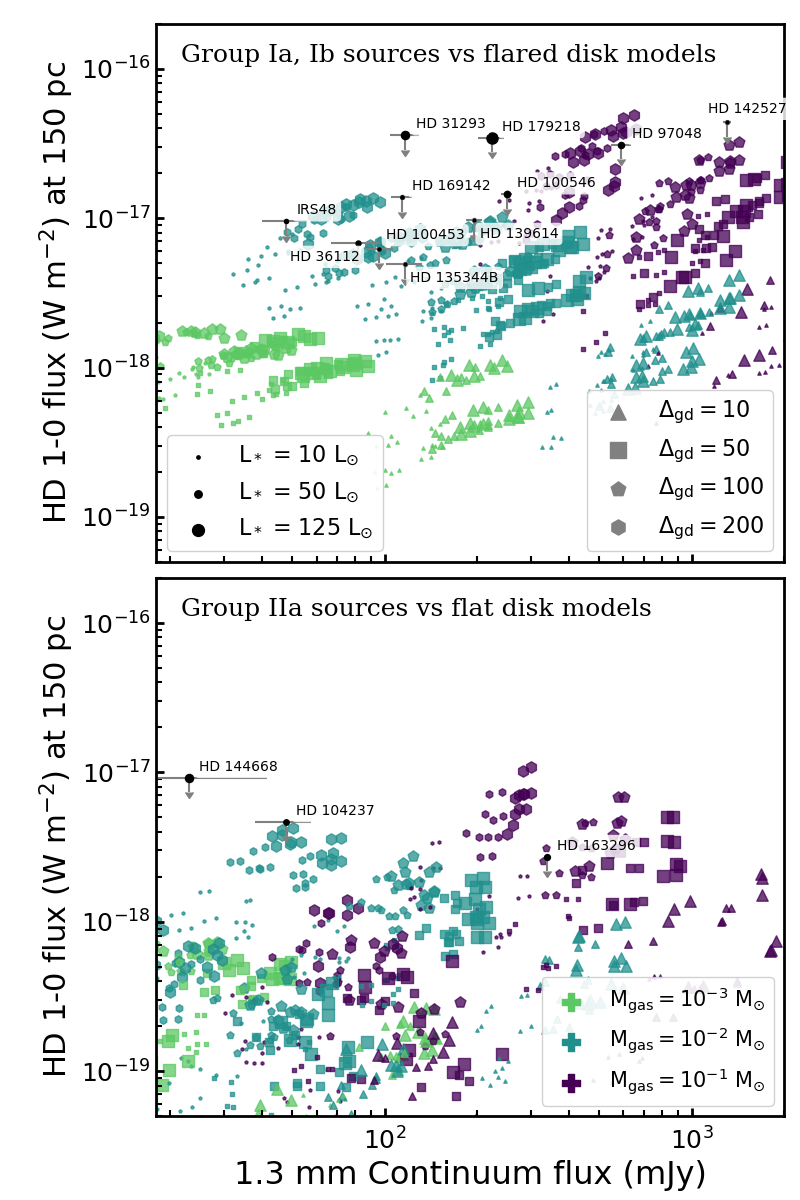}
\caption{HD 1\,-\,0 line flux versus 1.3 continuum fluxes for both observations and models. Panels shown here are similar to right panels of Figure \ref{fig:HD10_Mgas_1mm}, but also showing the model gas-to-dust mass ratios (marker shape) and stellar luminosities (marker size). }
\label{fig:hdfluxes 1-0 extended version}
\end{figure}

\end{appendix}

\end{document}